\documentclass[10pt,a4paper]{article}
\usepackage{amsmath,amssymb,latexsym}

\def\AFOUR{%
\setlength{\textheight}{8.5in}%
\setlength{\textwidth}{5.75in}%
\setlength{\topmargin}{-0.375in}%
\hoffset=-.5in%
\renewcommand{\baselinestretch}{1.17}%
\setlength{\parskip}{6pt plus 2pt}%
}

\AFOUR                                           

\makeatletter
\def\section{\@startsection {section}{1}{\z@}{-3.5ex plus -1ex minus
 -.2ex}{2.3ex plus .2ex}{\large\sc}}
\def\subsection{\@startsection{subsection}{2}{\z@}{-3.25ex plus -1ex minus
 -.2ex}{1.5ex plus .2ex}{\normalsize\sc}}
\makeatother

\makeatletter
\@addtoreset{equation}{section}

\makeatother

\newcommand{\nc}{\newcommand}
\newcommand{\rnc}{\renewcommand}

\nc{\be}{\begin{equation}}
\nc{\ee}{\end{equation}}
\nc{\bea}{\begin{eqnarray}}
\nc{\eea}{\end{eqnarray}}

\nc{\trac}[2]{{\textstyle\frac{#1}{#2}}}

\nc{\ex}[1]{\mbox{e}^{\,\textstyle#1}}

\nc{\CC}{\mathbb{C}}
\nc{\HH}{\mathbb{H}}
\nc{\PP}{\mathbb{P}}
\nc{\RR}{\mathbb{R}}
\nc{\ZZ}{\mathbb{Z}}
\nc{\II}{\mathbb{I}}
\nc{\EE}{\mathbb{E}}
\nc{\TT}{\mathbb{T}}
\nc{\DD}{\mathrm{I}\!\mathrm{D}}

\rnc{\d}{\delta}

\nc{\symx}{\circledS}
\nc{\ad}{\mathop{\mbox{ad}}\nolimits}
\nc{\tr}{\mathop{\mbox{tr}}\nolimits}
\nc{\Tr}{\mathop{\mbox{Tr}}\nolimits}
\nc{\Det}{\mathop{\mbox{Det}}\nolimits}
\rnc{\det}{\mathop{\mbox{det}}\nolimits}
\nc{\rk}{\mathop{\mbox{rk}}\nolimits}
\nc{\del}{\partial}
\nc{\diag}{\mathop{\mbox{diag}}\nolimits}
\nc{\ra}{\rightarrow}
\nc{\Ra}{\Rightarrow}
\nc{\LRa}{\Leftrightarrow}
\nc{\lra}{\leftrightarrow}
\nc{\ot}{\otimes}
\rnc{\ss}{\subset}
\nc{\nul}{\noindent\underline}
\nc{\non}{\nonumber\\}
\nc{\mat}[4]{\left(\begin{array}{cc}#1&#2\\#3&#4\end{array}\right)}
\rnc{\lg}{\frak{g}}
\nc{\G}[3]{\Gamma^{#1}_{\;{#2}{#3}}}
\nc{\nam}{\nabla_{\mu}}
\nc{\nan}{\nabla_{\nu}}
\nc{\dx}{\dot{x}}
\nc{\dxl}{\dot{x}^{\la}}
\nc{\dxm}{\dot{x}^{\mu}}
\nc{\dxn}{\dot{x}^{\nu}}
\nc{\ddx}{\ddot{x}}
\nc{\ddxm}{\ddot{x}^{\mu}}
\nc{\ddxn}{\ddot{x}^{\nu}}
\nc{\dxi}{\dot{\xi}}
\nc{\ddxi}{\ddot{\xi}}
\nc{\lsf}{\ell_s^{\mathrm{eff}}}
\nc{\lpf}{\ell_p^{\mathrm{eff}}}
\nc{\sqg}{\sqrt{g^{11}}}

\begin{document}


\vspace*{2cm}
\begin{center}
{\Large\sc Geometry of Schr\"odinger Space-Times II:\\[.2in] 
Particle and Field Probes of the Causal Structure}
\end{center}
\vspace{1cm}

\begin{center}
{\large\sc Matthias Blau, Jelle Hartong, Blaise Rollier}\\[.3cm] 
{Albert Einstein Center for Fundamental Physics\\
Institute for Theoretical Physics, Bern University\\ 
Sidlerstrasse 5,  CH-3012 Bern, Switzerland}
\end{center}

\vspace{1cm}

We continue our study of the global properties of the $z=2$ Schr\"odinger
space-time. In particular, we provide a codimension 2 isometric embedding
which naturally gives rise to the previously introduced global coordinates. 
Furthermore, we study the causal structure by probing the space-time with
point particles as well as with scalar fields. We show that, even though
there is no global time function in the technical sense (Schr\"odinger
space-time being non-distinguishing), the time coordinate
of the global Schr\"odinger coordinate system is, in a precise way,
the closest one can get to having such a time function. 
In spite of this and the corresponding strongly Galilean and almost
pathological causal structure of this space-time, it is nevertheless
possible to define a Hilbert space of normalisable scalar modes with
a well-defined time-evolution. We also discuss how the Galilean causal
structure is reflected and encoded in the scalar Wightman functions and
the bulk-to-bulk propagator.

 
\newpage

\begin{small}
\tableofcontents
\end{small}

\newpage

\section{Introduction}

Recently, following \cite{son,balamac,hrr,mmt,abmac,lif}, 
non-relativistic 
variants of the AdS/CFT correspondence have attracted considerable
attention.
This has brought to prominence deformations of (asymptotically)
AdS space-time geometries that exhibit (asymptotic) isometry groups which
are suitable Galilean counterparts of the relativistic conformal group,
such as the Schr\"odinger group.
These space-time geometries are interesting for at least three reasons:
\begin{enumerate}
\item First of all, of course, they are candidate gravitational duals
to non-relativistic strongly coupled (scale or conformally invariant)
condensed matter and other physical systems (for reviews see e.g.\ 
\cite{hartnoll} and \cite{mcgrev}).
This has led to new ways
of looking at well-known (if not well understood) physical phenomena,
but concrete and quantitative progress along these lines is currently
hampered by the lack of precise dual pairs, and by the fact that the
holographic dictionary in these space-times is still not nearly as well
understood as in the AdS case.

\item Secondly, and related to the issue just raised, this set-up
potentially provides one with a novel implementation of holography
which requires one to suitably generalise and modify the standard
AdS/CFT procedure.  While holography is (on fairly general and convincing
grounds) expected to be a generic feature of a quantum theory of gravity,
currently the only case that is reasonably well understood is that of
asymptotically AdS space-times. Attempts to generalise this to the
assymptotically flat or dS situation are fraught with technical and
conceptual complications. On the other hand, encouragingly some of the
AdS/CFT recipes \textit{do} appear to ``carry over'' in a simple-minded
way to the Schr\"odinger case. What is required now is a more systematic
understanding and underpinning of the calculational procedures, analogous
to that for AdS based on a suitable notion of conformal boundary, the
Fefferman-Graham expansion, and holographic renormalisation (for some 
concrete work along these lines see e.g.\ \cite{marika,ross}).

\item As a precursor
to this, one needs to gain an as precise understanding as possible of the 
properties of the model Schr\"odinger space-time that are shared with AdS,
and those that set it apart from AdS \cite{bhr}.
In particular, thirdly, Schr\"odinger space-time provides us with an interesting and
physically well-motivated example of a relativistic space-time that
exhibits a rather peculiar (and almost pathological) causal structure,
quite different from that of AdS (whose lack of global hyperbolicity is
its only mild, and well understood, potential source of pathology).
It is thus of interest, both in its own right and for the reasons mentioned
above, to study to which extent the behaviour of scalar fields, say, on such a space-time
is sensitive to, or  reflects, the Galilean oddities of this causal structure (as conventionally defined 
in terms of point particle probes and concepts).\footnote{This is similar in spirit
to the question to what extent (quantum) fields are sensitive to point particle
notions of singularities, see e.g.\ \cite{hm,bfw}.}
\end{enumerate}

In this paper we will discuss in some detail the issues raised in (3.) in the case of
the Schr\"odinger space-time with critical exponent $z=2$. In Poincar\'e-like coordinates
its metric takes the form
\be
\label{im}
ds^2 = -\beta^2 \frac{dt^2}{r^{4}} + \frac{1}{r^2}\left(-2dtd\xi + dr^2 + d\vec{x}^2\right)
\ee
(reducing to the AdS metric in Poincar\'e coordinates for $\beta=0$). This coordinate
system is geodesically incomplete (particles can reach $r=\infty$ in finite proper time
without encountering a singularity) \cite{bhr}. While some causal properties can be
(and have been \cite{hubeny,hrr}) reliably read off from the Poincar\'e patch metric \eqref{im}, 
a more detailed understanding of the global and causal properties requires a 
more global presentation and picture of the Schr\"odinger space-time. This is provided 
by the global coordinates introduced in \cite{bhr} in which the metric takes the form
\be
\label{igm}
ds^2 =-\beta^2 \frac{dT^2}{R^4} + \frac{1}{R^2}(-2dT dV -\omega^2(R^2 + \vec{X}^2)dT^2 + dR^2 +
d\vec{X}^2)\;\;.
\ee
This reduces to \eqref{im} for $\omega=0$, and also gives a (somewhat unusual) global coordinatisation
of AdS for $\beta = 0, \omega \neq 0$ (plane wave AdS). 
As a consequence, this allows us to directly compare and contrast
certain global properties of AdS (and of scalar propagation in this background) with those of the
Schr\"odinger space-time. 

To set the stage, in section 2 we discuss various aspects of Schr\"odinger geometry related to the 
global coordinates \eqref{igm}. In particular, in section 2.2 we provide a codimension 2 isometric
embedding of the Schr\"odinger space-time which naturally gives rise to these global cooordinates.
This embedding turns out to be not equivariant (i.e.\ not all isometries
are introduced from the isometries of the flat embedding space-time), and in appendix A we prove, 
using some group theory arguments, that indeed there are no codimension 2 equivariant isometric 
embeddings of the Schr\"odinger space-time.

In section 3 we study various aspects of the causal structure of
Schr\"odinger space-time. In section 3.1 we focus on those properties
that are common to the Schr\"odinger and plane wave AdS geometries.
Primarily these are properties of the
global time-coordinate $T$ of \eqref{igm}. In particular, we highlight the
fact that $T$, in spite of being a global function with $\del_T$ everywhere
timelike, is \textit{not} a time function in the strict sense. The
difference between AdS and Schr\"odinger is that the former is stably
causal and has a time function (the global time coordinate $\tau$
of the usual global AdS coordinates, for instance) while Schr\"odinger
is not stably causal and hence admits no time function whatsoever.
In this sense, $T$ turns out to be the closest one can get to having
a time function because it only fails to label causally related events
that lie on a $T=\text{cst}$ slice $\Sigma_T$. In section 3.2 we discuss
those aspects of the causal structure that are peculiar to $\beta \neq
0$, in particular the non-distinguishing character of this space-time
already noted in \cite{hubeny,hrr}, and the ensuing strongly Galilean character
of its causal structure.

Among the myriad of definitions and concepts related to the causal
properties of a space-time (see e.g.\ \cite{HE,Minguzzi:2006sa})
we have chosen to
focus on those aspects of the causal structure that we found to have
some counterpart in the analysis of scalar fields in the subsequent
section 4.  Here we will in particular address the question to which
extent time-evolution of a scalar field is affected by the absence of a
global time-function, and to which extent the Galilean causal structure
of the space-time is reflected and encoded in the Green's functions and
propagators of the theory. Some technical details have been relegated
to the appendices.

\section{Global Schr\"odinger geometry}

In this section, we briefly recall some basic features of the geometry of Schr\"odinger 
space-times and record some obervations regarding timelike Killing vectors. We also 
introduce
a codimension 2 isometric flat space embedding and discuss some aspects of Schr\"odinger
geometry that are particularly transparent from this embedding point of view, in particular
global coordinates.
 
\subsection{Isometries, timelike Killing vectors and global coordinates}\label{subsec:isometries}

The metric 
\be
\label{metrics}
ds^2 = -\beta^2 \frac{dt^2}{r^{4}} + \frac{1}{r^2}\left(-2dtd\xi + dr^2 + d\vec{x}^2\right)
\ee
($d\vec{x}^2 = (dx^1)^2 + \ldots (dx^d)^2$) is that of the
$(d+3)$-dimensional $z=2$ Schr\"odinger space-time $\mathsf{Sch}_{d+3}$
in Poincar\'e-like coordinates for $\beta^2 > 0$ and that of
$\mathsf{AdS}_{d+3}$ in null Poincar\'e coordinates for $\beta=0$.

For $\beta \neq 0$ this has the characteristic transitive Schr\"odinger
isometry algebra $\mathfrak{sch}(d)$ consisting of spatial rotations $M_{ab}$ and
translations $P_a$ and Galilean boosts $V_a$ (which we will not
make use of explicitly in the following, for details see appendix
A) and a central element $N=\del_{\xi}$ of null translations, as
well as of an $\mathfrak{sl}(2,\RR)$ subalgebra formed by
time-translations $H$, anisotropic dilatations $D$ and special
conformal transformations $C$,
\be
\begin{aligned}
H=\del_t \qquad D = 2t\del_t + r\del_r + x^a\del_a \qquad C= t^2\del_t + tr \del_r + tx^a \del_a +
\trac{1}{2}(r^2 + \vec{x}^2) \del_\xi\;\;.
\end{aligned}
\ee
In particular, the algebra
generated by the $\mathfrak{so}(d)$-singlets $\{H,C,D,N\}$, i.e.\ the 
isometry algebra of 3-dimensional Schr\"odinger spacetime,
is isomorphic to
\be
\mathfrak{sch}(d=0)\cong \mathfrak{so}(2,1)\oplus \RR_N\;\;.
\ee
For $\beta=0$ the Schr\"odinger isometry algebra  
$\mathfrak{sch}(d)$ is enhanced to the AdS isometry algebra $\mathfrak{so}(2,d+2)$, 
with $\dim \mathfrak{so}(2,d+2)- \dim\mathfrak{sch}(d) = 2(d+1)$.

The above Poincar\'e coordinate system is geodesically incomplete
as $r\ra\infty$ (geodesics can reach $r=\infty$ in finite affine
parameter, and the geometry is non-singular there) \cite{bhr}.  This
points to the necessity of introducing coordinates that also cover
the region beyond the Poincar\'e coordinate patch. A hint as to go about this
for $\beta \neq 0$ 
comes from analysing the timelike Killing vectors of this metric. 
For instance, the Killing vector $H=\del_t$ becomes lightlike at the Poincar\'e
horizon $r=\infty$, and is therefore not a suitable candidate for a global 
definition of time. If one considers, more generally, the 
linear combination
\begin{equation}
\label{ht}
\tilde{H} = a_HH + a_CC + a_NN +a_DD
\end{equation}
(these are the only relevant Killing vectors for these purposes), and calculates its norm
in Poincar\'e coordinates, one finds (for simplicity in the 3-dimensional case $d=0$ since
nothing essential changes for $d>0$)
\be
\label{htn}
\vert\vert\tilde{H}\vert\vert^2 = -\frac{\beta^2}{r^4}(a_H+2a_Dt + a_Ct^2)^2 
- \frac{2a_N(a_H+2a_Dt + a_Ct^2)}{r^2} + 
a_{D}^2-a_Ha_C\;\;.
\ee
Thus a necessary condition for $\tilde{H}$ to be timelike beyond the Poincar\'e horizon
is $a_{D}^2-a_Ha_C > 0$. The choice made in \cite{bhr} based on these and other considerations
was $\tilde{H} = H + \omega^2 C$. Introducing coordinates $(T,V)$ adapted to $\tilde{H}$ and 
the central element $N$, 
\be
\tilde{H}=\del_T  = H + \omega^2 C \quad,\quad N = \del_V\;\;,
\ee
the global metric reads
\be
\label{igma}
ds^2 =-\beta^2 \frac{dT^2}{R^4} + \frac{1}{R^2}(-2dT dV -\omega^2(R^2 + \vec{X}^2)dT^2 + dR^2 +
d\vec{X}^2)\;\;.
\ee
This coordinate system, in which the metric simply has the form of
a plane wave deformation of the Poincar\'e-like metric \eqref{metrics},
is geodesically complete for $\omega>0$ and reduces to the incomplete
Poincar\'e-patch metric for $\omega=0$.  In \cite{bhr} it was
moreover shown that this metric is closely related to the harmonic
trapping of non-relativstic CFTs that plays an important role in
the non-relativistic operator-state correspondence \cite{nison}.
In particular, time evolution with respect to the global time $T$
($\del_T$ is everywhere timelike)
is time-evolution with respect to the trapped Hamiltonian $H+\omega^2
C$, and the harmonic oscillator potential in the metric corresponds
to the trapping potential of the boundary theory.

One could, without loss of generality, choose $\omega=\beta=1$ for
the global metric, but we will keep $\omega$ and $\beta$ explicit
in order to facilitate the comparison of the properties of the
global Schr\"odinger metric with those of the Poincar\'e patch
metric and, in particular, with 
those of the AdS metric in global plane wave coordinates
(plane wave AdS) \cite{cgr,mmt,bhr} one obtains for $\beta=0$,
\be
\label{pwads}
ds^2 =\frac{1}{R^2}(-2dT dV -\omega^2(R^2 + \vec{X}^2)dT^2 + dR^2 + d\vec{X}^2)\;\;.
\ee
One other thing that one can read off and learn from \eqref{htn} is that the 
metric \eqref{metrics} for $\beta^2 < 0$ has no timelike Killing vectors for
$r\ra 0$ (since the first, now positive, term will dominate as $r\ra 0$). As
a consequence, even though the $\beta^2<0$ metric has Schr\"odinger isometry, 
it is not isometric to any patch of the global Schr\"odinger metric \eqref{igma}
(which has an everywhere timelike Killing vector). 
This illustrates that geometries
with Schr\"odinger isometry are not locally unique. 

In Poincar\'e coordinates and in global coordinates 
the metric is stationary (time-independent) but not static and one may
wonder whether there is (at least locally) any 
timelike Killing vector that is hypersurface-orthogonal. 
To analyse this question, let us once again consider the
linear combinations $\tilde{H}$ \eqref{ht}.
Imposing the integrability condition $\tilde{H}_{[\mu}\nabla_\nu \tilde{H}_{\rho ]} =0$,
one finds 
\begin{equation}
-\frac{\beta^2}{r^4}(a_H+2a_Dt + a_Ct^2)^2 
+(a_Ha_C -a_{D}^2)=0\;\;.
\end{equation}
For $\beta \neq 0$ the only solution is
$a_C=a_D=a_H=0$ so that $\tilde{H}\sim N$ which is not timelike but null (and hypersurface orthogonal
to the null surfaces $t=\mathrm{const.}$). An analysis in global coordinates leads to exactly
the same result, and we can
conclude that Schr\"odinger space-times are globally stationary 
but admit no static patch. For $\beta = 0$, on the other hand, one only finds the constraint 
$a_Ha_C-a_{D}^2=0$. 
A typical time-like solution is $a_C=a_D=0$ and $\tilde{H}=H+N$ which corresponds to choosing $x^0=t+\xi$ as the
new (and standard and obviously hypersurface-orthogonal) Poincar\'e time-coordinate. Of course, for
$\beta=0$ there are other Killing vectors, and global (and also hypersurface orthogonal) time $\tau$
in the usual global coordinates for AdS corresponds to the linear combination $\del_{\tau}=P_0+K_0$, 
$K_0$ a special conformal transformation.

\subsection{Isometric embeddings and global coordinates}

For the AdS space-time $\mathsf{AdS}_{d+3}$ there exists a codimension
1 isometric embedding into the pseudo-Euclidean space $\RR^{2,d+2}$.
It is relatively easy to see that no such embedding exists for
$\mathsf{Sch}_{d+3}$, more specifically that any hypersurface with
Schr\"odinger isometry is actually AdS-invariant.
Moreover, similar arguments show that there are no codimension 2 
equivariant isometric embeddings, i.e.\ isometric embeddings for which 
all the isometries are induced by isometries of the flat embedding 
space. We will establish these results in appendix A. 

However, a codimension 2 isometric (but not equivariant) embedding 
into $\RR^{2,d+3}$ equipped with the metric 
\be
ds^2 = -(dX^0)^2 + (dX^1)^2 + \ldots + (dX^{d+2})^2 - (dX^{d+3})^2 + (dX^{d+4})^2
\ee
exists and 
is given by
\be
\label{eq:pullback}
\begin{aligned}
(X^0,X^1) &= \frac{\xi\pm\frac{t}{2}+\frac{t}{2}\beta^2 f(t,r)}{r}\\
(X^{d+2},X^{d+3}) &= \tfrac{1}{2r}\left[\pm 1+2\xi t-\vec x^2-r^2-\beta^2 f(t,r)\right] \\
X^{1+a} &= \frac{x^a}{r} \qquad X^{d+4} = \tfrac{\sqrt{3}}{2}\beta f(t,r)
\end{aligned}
\ee
where $a=1,\ldots,d$ and where $f(t,r) =\frac{t^2+1}{r^2}$. 
Indeed, the metric induced by this embedding on the  codimension 2 surface parametrised by
$(t,\xi,r,\vec{x})$ is precisely the Schr\"odinger/AdS metric in Poincar\'e coordinates
\eqref{metrics}.
Explicitly, the $(d+5)$ coordinates are related by the two constraints
\be
\label{constraints}
\begin{aligned}
-(X^0)^2+(X^1)^2+\sum_a (X^{1+a})^2+(X^{d+2})^2-(X^{d+3})^2 &=-1 - \tfrac{4}{3}(X^{d+4})^2 \\
\beta\left[\left(X^0-X^1\right)^2+(X^{d+2}-X^{d+3})^2\right]&=\tfrac{2}{\sqrt{3}}X^{d+4}
\end{aligned}
\ee
and the inverse transformation, subject to these constraints, is
\be
\label{inverse}
\begin{aligned}
t &=  \frac{X^0-X^1}{X^{d+2}-X^{d+3}} \qquad 
r =  \frac{1}{X^{d+2}-X^{d+3}} \qquad
x^a = \frac{X^{1+a}}{X^{d+2}-X^{d+3}}\\
\xi &=  \frac{1}{2}\left[\left(\frac{X^0+X^1}{X^{d+2}-X^{d+3}}\right)-
\frac{2\beta X^{d+4}}{\sqrt{3}}\left(\frac{X^0-X^1}{X^{d+2}-X^{d+3}}\right) \right]
\end{aligned}
\ee
The parameter $\beta$ describes the deformation away from $AdS_{d+3}$ and
for $\beta=0$ one reproduces the standard codimension 1 embedding into
the hyperplane $\RR^{2,d+2}\subset \RR^{2,d+3}$ given by $X^{d+4}=0$.
For $\beta \neq 0$, $X^{d+4}$ is non-trivial and the first constraint
equation describes a surface that can be viewed as $AdS_{d+3}$ space-time
of variable AdS radius where the radius is a function of $X^{d+4}$.
Just as for AdS, in order not to have closed time-like curves we work
with the universal cover.

As already alluded to above, the above isometric embedding
has the property that not all the Schr\"odinger
isometries are actually induced by the $ISO(2,d+3)$-isometries of
the embedding space $\RR^{2,d+3}$. Indeed, for $\beta \neq 0$ the
isometries that embed into $SO(2,d+3)$ (all the translational symmetries
are manifestly broken by the constraints) are those of the constant
$X^{d+4}$ slices, namely $M_{ab},P_a,V_a,N$ and $H+C$, while the 
``accidental'' additional isometries are $D$ and $H-C$.  For instance, a
shift in $t$ (generated by $H$) is induced by a non-linear transformation
of the coordinates $X^A$ for $\beta \neq 0$ whereas it is realised by a
linear $SO(2,2)$-transformation in the $(X^0,X^1,X^{d+2},X^{d+3})$-plane
for $\beta =0$, as it should be.  

The geodesic distance between two points (relevant for our discussion
of scalar fields and Green's functions in section 4) is invariant under
the (simultaneous) action of the isometry group of a space-time on the
two points.  If we had an equivariant isometric embedding, we could
introduce at least one isometry-invariant notion of the distance between
two points in terms of the standard pseudo-Euclidean distance between two
points in the embedding space.  In the AdS case $\beta =0$ this gives
rise to the usual chordal distance and its relation with the geodesic
distance. For $\beta \neq 0$, however, this option is not available (the
induced distance function is not a Schr\"odinger invariant object). We
will discuss and construct these invariants (it turns out that there
are two independent such functions) in appendix B.

In spite of its shortcomings, the above embedding is quite useful for a number of things. For
instance, the constraints \eqref{constraints} suggest a natural parametrisation 
of the form
\be
\begin{array}{lll}
X^0-X^1 = \tfrac{\sin T}{R} &\qquad&  X^0+X^1 = \tfrac{1}{R}\left(2V\cos T+b\sin T\right) \\
X^{d+2}-X^{d+3} = \tfrac{\cos T}{R} && X^{d+2}+X^{d+3}  = \tfrac{1}{R}\left(2V\sin T-b\cos T\right) \\
X^{1+a} = \tfrac{X^a}{R} && X^{d+4} = \tfrac{\beta\sqrt{3}}{2R^2} 
\end{array}
\ee
with which the first constraint reduces to $b = R^2 + \vec X^2+\tfrac{\beta^2}{R^2}$.
Then the induced metric is
precisely the $\omega = 1$ case of the global plane wave Schr\"odinger metric
\eqref{igma}. From the present embedding point of view we learn that this
parametrisation indeed covers the entire space-time (both for the codimension 1 embedding of AdS
for $\beta=0$ and for $\beta \neq 0$), and that what
was a geodesically complete coordinate system in \cite{bhr} is
now also global from the embedding point of view. It follows from 
\eqref{inverse} that the Poincar\'e patch only covers the region $X^{d+2}-X^{d+3}>0$.
This isometric embedding generalises the embedding of plane waves
found  a long time ago in \cite{Rosen,Collinson}  (see also \cite{bfp}),
and correspondingly the equivariantly realised isometries $M_{ab},P_a,V_a,N$
and $H+C$ form the isometry algebra of an isotropic symmetric plane
wave \cite{bfp,mm}.

Another issue that is particularly transparent from the embedding
point of view is that of potential conical singularities that arise
if one compactifies the $V$ (or, equivalently, $\xi$) direction
\cite{mmt,abmac}. The situation turns out to be identical for AdS
and Schr\"odinger.  First of all we note that the shift $V\rightarrow
V+\alpha$ is a symmetry of both the two constraint equations as well as
of the embedding space-time for any $\alpha\in\RR$. We want to see what
happens if we identify $V\sim V + 2\pi L$. 

Using that
\begin{equation}
V = \frac{1}{2}\left( \frac{(X^0-X^1)(X^{d+2}+X^{d+3})+(X^0+X^1)(X^{d+2}-X^{d+3})}{(X^0-X^1)^2+(X^{d+2}-X^{d+3})^2}\right)
\end{equation}
we see that the identification of $V$ with $V+ 2\pi L$ leads to the identifications
\be
\begin{aligned}
X^0+X^1 &\sim X^0+X^1+ 4\pi L(X^{d+2}-X^{d+3})\\
X^{d+2}+X^{d+3} &\sim X^{d+2}+X^{d+3}+ 4\pi L(X^0-X^1)\;\;.
\end{aligned}
\ee
We therefore conclude that there are two conical singularities:
\begin{enumerate}
\item At $(X^0+X^1,X^{d+2}-X^{d+3})=(0,0)$ which can be reached in the limit $R\rightarrow\infty$ with $\sin T=0$ fixed (for $\omega=1$) and $V,X$ finite but arbitrary.
\item At $(X^{d+2}+X^{d+3},X^0-X^1)=(0,0)$ which can be reached in the limit $R\rightarrow\infty$ with $\cos T=0$ fixed (for $\omega=1$) and $V,X$ finite but arbitrary.
\end{enumerate}
In Poincar\'e coordinates the limit mentioned in point 1 corresponds
to the limits $r\rightarrow\infty$, $t/r\rightarrow 0$ and $\vec
x/r\rightarrow 0$. This is in agreement with the comments made
in \cite{abmac} regarding the conical singularity after
compactification of $\xi$. The singular locus of point 2, on the other hand, 
lies outside the Poincar\'e patch.

\section{Point particle probes of the causal structure}

In this section we will discuss the causal structure associated with
point particles moving along causal curves in the space-time with
global metric \eqref{igma},
\begin{equation}\label{globalmetric}
ds^2=-\left(\frac{\beta^2}{R^4}+\frac{\omega^2}{R^2}(R^2+\vec
X^2)\right)dT^2+\frac{1}{R^2}\left(-2dTdV+d\vec X^2+dR^2\right)\,.
\end{equation}
We will start with $\beta$-independent properties, i.e.\ those
that 
also hold in the geodesically complete plane wave AdS space-time. We then
explore causality statements which are specific to the Schr\"odinger
space-time.  We will focus on those aspects of the
causal structure that are relevant to our analysis of
scalar fields in section 4. Definitions follow the standard reference
\cite{HE} and the more recent review \cite{Minguzzi:2006sa}.

\subsection{Time functions and time coordinates}

First of all, let us collect some basic properties of the global
coordinate $T$:
\begin{enumerate}
\item $T$ is a globally defined smooth function.
\item The vector field $\partial_T$ is an everywhere timelike Killing vector. In particular, 
it provides a time orientation.
\item The gradient of $T$ is null.
\item $T$ is strictly increasing along any future-directed timelike
curve.
\item $T$ is non-decreasing along any future-directed null curve.
\end{enumerate}
The first two points are trivial and follow from the fact that
(\ref{globalmetric}) is a global coordinate system. The third follows 
from $g_{VV}=0$. The last two
points can be summarised saying that $\dot{T}\geq 0$ along any 
future-directed causal curve, with $\dot{T}>0$ for timelike curves, implying
that the space-time is chronological (no closed timelike curves can
occur). This can be seen as follows: take any curve $\gamma(\lambda)$
with tangent $\left(\dot T,\dot V,\dot R,\dot{\vec X}\right)$ and
require it to be causal,
\begin{equation}\label{causalconstraint}
\left( \frac{\beta^2}{R^2} + \omega^2(R^2 +\vec X^2) \right)\dot{T}^2
+ 2\dot{T}\dot{V} \geq \dot{R}^2+{\dot{\vec X}}{}^2\,,
\end{equation}
and future-directed\footnote{A curve $\gamma$ is future-directed with
respect to a timelike vector field $X^\mu$ if
$g_{\mu\nu}X^\mu\dot{\gamma}^\nu<0$.} with respect to the timelike
vector field $\left(\tfrac{\partial}{\partial T}\right)^\mu$,
\begin{equation}\label{futuredirected}
\left(\frac{\beta^2}{R^4}+\frac{\omega^2}{R^2}(R^2+\vec
X^2)\right)\dot T+\frac{\dot V}{R^2}>0\,.
\end{equation}
Then, since the right hand side of equation \eqref{causalconstraint}
is greater than or equal to zero it follows that
\begin{equation}\label{causality2}
\dot T\left(\left(\frac{\beta^2}{R^2}+\omega^2(R^2+\vec
X^2)\right)\dot T+\dot V\right)\ge -\dot T\dot V\,.
\end{equation}
Now we prove that $\dot T\ge 0$ by arguing that $\dot{T}<0$ leads to a
contradiction. Suppose $\dot T<0$, then equations
\eqref{futuredirected} and \eqref{causality2} implies $\dot T\dot V>
0$ but since $\dot T<0$ it must be that $\dot V< 0$ which is then in
contradiction with equation \eqref{futuredirected}. Hence, we must
have $\dot T\ge 0$ along all future-directed causal curves. Similarly,
by restricting (\ref{causalconstraint}) to timelike curves one shows
that $\dot T>0$ along all future-directed timelike curves (statement
4). Furthermore, one observes from (\ref{futuredirected}) that if
$\dot{T}=0$ then one necessarily has $\dot{V}>0$ so that no closed
causal curve can ever be formed for non-compact $V$. This shows that
the space-time is causal. In the compact $V$ case, closed causal curves
exist (by construction), and the space-time is only chronological.

A time function is a globally defined continuous function that is strictly
increasing along all future-directed causal curves. It therefore provides
an ordering, as all causally related events can then be labeled by
different values of $T$. The existence of a time-function is equivalent to
the space-time being stably causal, and this in turn is equivalent to the
existence of a (not necessarily the same) globally defined function whose
gradient is everywhere timelike \cite{HE,Minguzzi:2006sa}.
Because there exist future-directed
causal curves for which $\dot{T}=0$, $T$ is not a time function (and neither
is the gradient of $T$ everywhere timelike; in fact, as mentioned above, it is
everywhere null). So what about stable causality of these space-times?

AdS is well known to be stably causal; thus
even though $T$ is not a time function one \textit{can} find a time
function for $\beta = 0$ (this global time function can be taken to be $\tau$, the time
coordinate of the usual global AdS coordinate system). But, as we will
see in section 3.2, the Schr\"odinger space-time ($\beta\neq 0$) is not stably
causal and hence it does not admit any time function. In that respect,
$T$ is the closest one can get to having a time function because it only
fails to distinguish causally related events that lie on a $T=\text{cst}$
slice $\Sigma_T$. 

Such causally related events with the same $T$ are related by so-called
lightlike lines.\footnote{A lightlike line is an achronal inextendible
causal curve \cite{Minguzzi:2008ch}.  A set $S$ is called achronal
resp.\ acausal if no two distinct points of $S$ can be connected by a
timelike resp. causal curve.} Indeed, a chronological space-time without
such lightlike lines would be stably causal \cite{Minguzzi:2008ch}.
Lightlike lines are always null geodesics but the converse is generally
not true. In space-times such as Minkowski and AdS all null
geodesics are lightlike lines, so that the existence of lightlike lines
alone does not signal any pathology. 

From what we proved so far
(statement 4) it follows that the surfaces $\Sigma_T$ are achronal
but not acausal. Hence in our context the lightlike lines are given by
\begin{equation}\label{nullVcurve}
\gamma(\lambda)=(T_0,V(\lambda),R_0,\vec X_0)\,,
\end{equation}
where $V(\lambda)$ is a strictly monotonically increasing function of
$\lambda$. These are precisely the null geodesics 
(affinely parametrised for $V(\lambda)\sim\lambda$) 
with zero lightcone momentum $P_V \equiv P_-=0$ (cf.\ appendix D).
The tangent is $u^\mu=(0,\dot V,0,\vec 0)$, 
so that we have $g_{\mu\nu}\left(\tfrac{\partial}{\partial
T}\right)^\mu u^\nu<0$ from which it follows that $\gamma$ is a
future-directed null geodesic along which the time coordinate $T$
remains constant. 

Finally, let us us note that,  as a consequences of the existence of these
lightlike lines, the future domain of dependence of a constant time
slice $\Sigma_T$, denoted by $D^+(\Sigma_T)$, is empty.\footnote{The 
future domain of dependence $D^+(S)$ of a
set $S$ is the set of points $p$ such that every
past inextendible causal curve through $p$ intersects $S$.} 
This has to be contrasted with AdS in the usual
global coordinates where the future domain of dependence of a global
time slice $\tau$ is not empty. 
Actually there are
two distinct sources for the emptiness of the future and
past domain of dependence: one has $D^\pm(\Sigma_{T_0}) = \emptyset$ because
\begin{enumerate}
\item for each time $T>T_0$ there exists a future and past
inextendible null geodesic that has $\dot T=0$, those are the 
lightlike lines \eqref{nullVcurve};
\item for any arbitrary point $P=(T_0\pm\delta, V_0, R_0, \vec X_0)$,
say, with $\delta>0$, that lies to the future (+) or past (-) of $T_0$
there exists a, respectively, past or future inextendible timelike
curve that goes all the way to the boundary at $R=0$ without crossing
the slice $\Sigma_{T_0}$. An example of such a timelike curve is given by
\begin{equation}
\gamma_{\textit{past}}(\lambda) = \left(\begin{array}{c}
T_0-\frac{\delta}{2}e^{-2\lambda}-\frac{\delta}{2} \\
\frac{R_0^2}{\delta}\lambda + V_0\\
R_0 e^{-\lambda} \\
\vec X_0
\end{array}\right) \qquad
\gamma_{\textit{future}}(\lambda) = \left(\begin{array}{c}
T_0+\frac{\delta}{2}e^{2\lambda}+\frac{\delta}{2} \\
\frac{R_0^2}{\delta}\lambda + V_0\\
R_0 e^{\lambda} \\
\vec X_0
\end{array}\right)
\end{equation}
\end{enumerate}

\subsection{Galilean-like causal structure}

The dramatic effect of having a non-zero $\beta$ is that it makes the
space-time non-distinguishing\footnote{A space-time is called
non-distinguishing if there exist two distinct points that have
identical past and future. Non-distinguishing space-times do not admit
any time function.} whereas it is stably causal for AdS. This has
already been proven in \cite{hubeny} for the $z=3$
Schr\"odinger space-time using the Poincar\'e patch (and the possible
connection of this property with a Galilean-like causal structure
was noted in \cite{hrr}). The proof is
based on the existence of a causal curve that connects any two points
whose time interval is infinitesimally small. It can be shown that
such a curve can be constructed for any $z>1$ and because there exists
a local defining property for a space-time to be distinguishing
\cite{Minguzzi:2006sa}, being non-distinguishing in the Poincar\'e
patch is enough to assure that the space is also non-distinguishing
globally.\footnote{Alternatively, one can prove that the $z>1$ Schr\"odinger space-times are
non-distinguishing by observing i) that in the Poincar\'e-like
coordinate system they are conformal to a class of pp-wave space-times
that in \cite{hubeny2} have been proven to be non-distinguishing, and ii) that
being non-distinguishing is a local property of a space-time which is
therefore preserved under conformal transformations.}
For $z=2$, such a curve can also be constructed directly in
global coordinates, leading to the same conclusions (see appendix C).

In appendix \ref{sec:chronologicalfuture}, we show explicitly that the
chronological future (past) $I^\pm(p_0)$ of any point $p_0=(T_0,V_0,R_0,\vec X_0)$ 
on the slice $\Sigma_{T_0}$ is the set of all points with $T>T_0$ ($T<T_0$).
Therefore, for any point $p_0$ one has the decomposition 
\be
\label{p0}
\mathsf{Sch}=I^-(p_0) \cup \Sigma_{T_0}\cup I^+(p_0)
\ee
of the Schr\"odinger space-time.
Since all points
on a constant time slice share the same future and past,
the space-time is in a sense ``maximally non-distinguishing''.   

This is strongly reminiscent of a Galilean causal structure and 
Galilean relativity. 
In order to sharpen this analogy, we need an
appropriate notion of spacelike separation. We will call 
two points $x$ and $x'$ \textit{spacelike separated} if there
is no causal curve connecting them. It is perhaps worth
pointing out that this notion of spacelike separation does not imply
that two points are spacelike separated when they 
can be connected by a spacelike geodesic: there are spacelike geodesics
along which $\dot T\neq 0$, while we already know that any two points with
$T\neq T'$ can be connected by a timelike curve. 
This means that spacelike separated points necessarily lie on an equal-time
slice $\Sigma_T$.

This appears to be completely Galilean, since in Galilean relativity
any two non-simultaneous events can be connected by the worldline
of a (sufficiently fast moving) particle, and the only events
for which no such curve exists are those that are simultaneous.
However, the novel and non-Galilean feature of the causal structure
of Schr\"odinger space-times is the presence of lightlike lines.
Indeed, on a Schr\"odinger space-time all points with the same value
of $T$ are either spacelike separated or separated by a lightlike
line and conversely all points that are either spacelike separated
or separated by a lightlike line lie on an equal time $T$ surface.
While any time coordinate on the Schr\"odinger space-time whose
values label the slices $\Sigma_T$ plays the role of some absolute
(Galilean) time, the null coordinate $V$ (affinely) parametrises
the lightlike lines and thus that part of the surfaces $\Sigma_T$
that has no Galilean counterpart.

This Galilean-like structure is preserved by the subgroup
\be
\label{fpd}
 (T',V',R',\vec{X}') =  (T'(T),V'(T,V,R, \vec X), R'(T,R, \vec X), \vec X'(T,R, \vec X))
\ee
of the full group of space-time diffeomorphisms.
Indeed, any set of
coordinates $(T',V',R',\vec X')$ obtained by acting on the global
coordinates $(T,V,R,\vec X)$ with such a diffeomorphism is such that
$T'$, the new time coordinate, labels surfaces of spacelike and
lightlike line separated events while any new $V'$ coordinate
parametrises the lightlike lines. The normal to a constant
$T'$ slice $\Sigma_{T'}$ is proportional to the null Killing vector
$N$, and the (degenerate)
induced metric on $\Sigma_{T'}$ agrees with the
Galilean metric measuring the distance between simultaneous
(spacelike) separated events. This special class of diffeomorphisms
consists precisely of the double foliation preserving diffeomorphisms
discussed in a related context in \cite{Horava:2009vy}.
Here the double foliation
refers to the foliations associated with the equal time surfaces
and the lightlike lines.

\section{Scalar field probes of the causal structure}

In this section we will study the causal structure of Schr\"odinger
space-times as seen by scalar field probes and show that, even though
the causal structure seen by point particles is close to pathological,
this is not so from the point of view of the scalars.

\subsection{Canonical analysis}\label{subsec:canonicalanalysis}

The action for a massive complex scalar field $\phi$ is
\begin{equation}\label{eq:action}
S=-\int
d^{d+3}x\sqrt{-g}
\left(\partial_\mu\phi^*\partial^\mu\phi+m_0^2\phi^*\phi\right)+\ldots\,,
\end{equation}
where $m_0$ is a mass parameter and the dots refer to intrinsic boundary terms, e.g. terms that only 
involve the scalar $\phi$, its tangential derivatives along the boundary and the induced boundary metric. 
We will consider scalar fields $\phi$
that are eigenstates of the central element $\del_V$ of the Schr\"odinger algebra, i.e.\
\begin{equation}\label{UIR}
\phi(T,V,R, \vec X) =e^{-imV}\psi(T,R, \vec X)\,,
\end{equation}
in which $m\neq 0$, 
and we will decompose solutions to the scalar field equation formally as
\begin{equation}\label{eq:formaldecomposition}
\phi=\sum_M a_M u_M\,,
\end{equation}
where the $u_M(T,V,R, \vec X)$ form a complete
set of modes with a fixed momentum $m$ in the $V$ direction,
$u_M(T,V,R, \vec X)=e^{-imV}v_M(T,R, \vec X)$.
These states furnish a unitary irreducible representation of
the Schr\"odinger group with respect to the inner product
\begin{equation}\label{innerproduct}
 \langle u_M\vert u_{M'}\rangle=\frac{i}{2}\int_{\Sigma_T}d\Sigma^\mu
u_M^*\overleftrightarrow{\partial_\mu}u_{M'}\,.
\end{equation}
The $T=\text{cst}$ slice $\Sigma_T$ is a
lightlike surface whose normal is $\left(\tfrac{\partial}{\partial
V}\right)^{\mu}=\delta^\mu_V$. The integration measure is
$d\Sigma^\mu=\delta^\mu_VR^{-(d+1)}dRd^d\vec XdV$.
Irreducibility follows from 
irreducibility with respect to the centrally extended
Galilean subgroup.

We denote the Killing vectors of the Schr\"odinger metric collectively
by $k_A= k_A^\mu\del_\mu$. From the Noether theorem one obtains the corresponding 
conserved currents
\begin{equation}
j^\mu_A=\sqrt{-g}\,k_A^\nu T^\mu{}_{\nu}\,,
\end{equation}
where $T_{\mu\nu}$ is the energy momentum tensor. 
We define the corresponding charges $K_A$ by
\begin{equation}\label{eq:charges}
K_A=\int_{\Sigma_T}dVd^d\vec XdR\,j_A^T=\int_{\Sigma_T}d\Sigma^\mu
k_A^\nu T_{\mu\nu}\,.
\end{equation}
For fields $\phi$ of the form $\phi=e^{-imV}\psi(T,R,\vec X)$ the
charges $K_A$ can be written as
\begin{equation}\label{charges}
K_A=\int_{\Sigma_T}dVd^d\vec XdR\left(\pi k_A\phi+\pi^*
k_A\phi^*-k_A^T\mathcal{L}\right)\,,
\end{equation}
where $k_A^T$ is the $T$ component of the Killing vector $k_A$,
$\mathcal{L}$ denotes the scalar field bulk Lagrangian and where $\pi$
denotes the canonical momentum
\begin{equation}
\pi=\tfrac{\partial\mathcal{L}}{\partial(\partial_T\phi)}=R^{-(d+1)}
\partial_V\phi^*\,.
\end{equation}
The canonical momentum is thus not independent of the initial data
$\phi^*(T,V,R, \vec X)$ specified at some equal time $T$ surface, and
imposing vanishing equal time Poisson brackets 
between
$\phi(T,V,R,\vec X)$ and $\phi^*(T,V',R',\vec X')$ would be inconsistent
with a non-vanishing Poisson bracket 
$\{\phi(T,V,R,\vec X),\pi(T,V',R',\vec X')\}\neq 0$.
This problem is resolved by taking the following
Poisson bracket (for fields with the same nonzero $m$):
\begin{equation}\label{eq:SchPoissonbracket}
\{\phi(T,V,R, \vec X),\phi^*(T,V',R, \vec
X')\}=f(V-V')R^{d+1}\delta(\vec X-\vec X')\delta(R-R')\,.
\end{equation}
The Poisson bracket for fields with different $m$ is taken to vanish.
The function $f(V-V')$ will be chosen such that
\begin{equation}\label{eq:actioncharges}
\{K_A,\phi\} = -k_A\phi\,,
\end{equation}
upon use of the Euler--Lagrange equations of the Lagrangian given in
\eqref{eq:action}. If we consider the Hamiltonian $H_T$ associated
with the Killing vector $\partial/\partial T$ then this requirement
means that the Hamilton equations and the Euler--Lagrange equations
coincide. In the definition of the charges $K_A$ \eqref{eq:charges},
the $dV$ integral ranges from $V_1$ to $V_2$ where $V_1\neq V_2$ are
arbitrary finite points.

There exists a unique function $f(V-V')$ which is such that
\eqref{eq:actioncharges} holds true for any choice of $V_1$ and $V_2$.
This function is given by
\begin{equation}\label{def1}
f(V-V')=\frac{-i}{2m(V_2-V_1)}e^{-im(V-V')}\,.
\end{equation}
When we compactify $V$ by identifying $V\sim V+2\pi L$ then we should
replace in the function $f$ the momentum $m$ by the discrete momentum
$m/L$ with $m\in\mathbb{Z}$ and write $V_2-V_1=2\pi L$, so that we get
\begin{equation}\label{def2}
f(V-V')=\frac{-i}{4\pi m}e^{-im(V-V')/L}\,.
\end{equation}
For the case of a Schr\"odinger space-time with a non-compact $V$ we
will from now on take $V_2-V_1=2\pi$. 
This choice will prove useful later on. It is the value for which
results obtained for the free scalar field on a Schr\"odinger
space-time after integration over $m$ gives us 
(for $\beta = 0$) the
corresponding result on plane wave AdS \eqref{pwads}. Also, for this value the
Poisson brackets for compact and non-compact $V$ are
identical.

The form of the Schr\"odinger Poisson bracket can also be understood
by starting with the Poisson bracket for scalar fields in plane wave
AdS and decomposing it into modes with a fixed momentum in the
$V$ direction. If we then fix the momentum $m$ the resulting Poisson
bracket takes the Schr\"odinger form. To see this consider plane wave
AdS with a compact $V$ coordinate. Because $T$ is like a lightcone
time coordinate we must once again define equal $T$ Poisson brackets
for $\phi(T,V,R,\vec X)$ and $\phi^*(T,V',R',\vec X')$. Borrowing from
the result used in lightcone quantisation in Minkowski space-time
with a compact null circle \cite{Heinzl:2000ht}, appropriately
generalised to AdS, the Poisson brackets turn out to also have the
form \eqref{eq:SchPoissonbracket}, with
\be
f(V-V') = -\tfrac{1}{2}\left(\tfrac{1}{2}\text{sign}(V-V')-\tfrac{V-V'}{2\pi
L}\right)\;\;.
\ee
The sign function can be decomposed into Fourier modes as
\begin{equation}\label{eq:decompositionsignfunction}
\tfrac{1}{2}\text{sign}(V-V')-\tfrac{V-V'}{2\pi L}=\sum_{m\neq
0}\tfrac{i}{2\pi m}
e^{-im(V-V')/L}\;\;,
\end{equation}
and substituting the corresponding mode decomposition
\begin{equation}\label{eq:AdScompactVdecomposition}
 \phi(T,V,R, \vec X)=\psi_0(T,R, \vec X)+\sum_{m\neq 0}\psi_m(T,R, \vec
X)e^{-imV/L}
\end{equation}
into the Poisson bracket, 
we find that the functions $\psi_{m\neq 0}$ satisfy the
Schr\"odinger Poisson bracket of \eqref{eq:SchPoissonbracket} with the
function $f$ precisely as in \eqref{def2}. 

As regards the $m=0$ modes, let us first note that 
they have an arbitrary time-dependence that is not fixed by the Klein--Gordon
equation. Since these are the modes with zero
lightcone momentum, $P_-\phi =0$, they can be thought of as the
precise scalar field counterparts of the lightlike lines discussed in section 3. 
It turns out that these modes vanish for plane
wave AdS with a compact $V$ coordinate and for a free non-interacting
theory (see \cite{Heinzl:2000ht} for an explanation of this fact in
Minkowski space-time with a compact null circle). This follows from
substituting the decomposition \eqref{eq:AdScompactVdecomposition} into
the Hamiltonian. One of Hamilton's equations is then the statement
that $\psi_0=0$. The problems encountered with the $m=0$ modes in
\cite{Hellerman:1997yu} appear when one studies loop corrections in
an interacting theory. This lies beyond the scope of our work and
it might be interesting to see what kind of interacting theories
on a Schr\"odinger space-time with a compact lightlike circle are
perturbatively well-defined.

To obtain the normalisable as well as the non-normalisable modes we
impose the condition that solutions are regular everywhere in the
bulk. The normalisable modes must furthermore satisfy the boundary
condition that the inner product \eqref{innerproduct} is time
independent. This will be the case provided we have
\begin{equation}\label{eq:noflux1}
 \lim_{\varepsilon\to
0}\int_{R=\varepsilon}R^{-(d+1)}u_M^*\overleftrightarrow{\partial_R}u_
{M'}dVd\vec X=0\,.
\end{equation}
This is the condition that the flux of the current
$u_M^*\overleftrightarrow{\partial_\mu}u_{M'}$ through the boundary at
$R=0$ vanishes. Imposing this boundary condition requires that $\nu$
defined by
\begin{equation}
 \nu=\sqrt{\tfrac{(d+2)^2}{4}+m_0^2+\beta^2\,m^2}\label{eq:nu}
\end{equation}
is real so that all normalisable modes respect the
Breitenlohner--Freedman bound \cite{Breitenlohner:1982jf}.
There are two set of modes compatible with this boundary condition.\footnote{These $\pm$ normalisable 
modes have also been
discussed in global coordinates in \cite{Moroz:2009kv} and in
Poincar\'e coordinates in \cite{son}.
For a different
class of solutions, with cut-off dependent boundary conditions allowing for
imaginary $\nu$ (violating the Breitenlohner-Freedman bound) 
see \cite{Moroz:2009kv}.} 
They are given by
\begin{eqnarray}
 \phi_\pm & = &
e^{-imV}\sum_{L,n,k}a^\pm_{L,n,k}v^\pm_{L,n,k}\nonumber\\
&=&e^{-imV}\sum_{L,n,k}C^\pm_{L,n,k}a^\pm_{L,n,k}e^{-iE^\pm_{L,n,k}T}
Y_Le^{-\tfrac{1}{2}\omega\vert m\vert(\rho^2+R^2)}\rho^L
R^{\Delta_\pm}\times\nonumber\\
&&\times L_n^{L-1+d/2}(\omega\vert
m\vert\rho^2)L^{\pm\nu}_k(\omega\vert m\vert
R^2)\,,\label{decomposition}
\end{eqnarray}
where
\begin{equation}
 \Delta_\pm=\frac{d+2}{2}\pm\nu\,.
\end{equation}
The energy of the $+/-$ modes is given by
\begin{equation}
E^\pm_{L,n,k}=\text{sign}(m)2\omega\left(n+k+\tfrac{L}{2}+\tfrac{
\Delta_\pm}{2}\right)\,,
\end{equation}
with $L,n,k=0,1,2,\ldots$. For the minus modes we must assume
$0<\nu<1$ while for the plus modes we must assume that $\nu>0$. The
cases $\nu=0,1,2,\ldots$ have to be dealt with separately because they
involve logarithmic solutions. Here we will always assume that
$\nu\neq 0,1,2,\ldots$.

The constant $C^\pm_{L,n,k}$ will be chosen such that upon
quantisation the creation and annihilation operators $a^\pm_{L,n,k}$
and ${a^\pm}^\dagger_{\!\!\!\!\!L,n,k}$ satisfy the commutation
relation
\begin{equation}\label{creationannihilationcommutator}
 [a^\pm_{L,n,k},{a^\pm}^\dagger_{\!\!\!\!\!L',n',k'}]=\tfrac{1}{2}
\text{sign}(m)\delta_{LL'}\delta_{nn'}\delta_{kk'}\,.
\end{equation}
The constant $C^\pm_{L,n,k}$ can be taken to be real and positive and
is found to be
\begin{equation}
 (C^\pm_{L,n,k})^2=\frac{2(\omega\vert m\vert)^{L+\Delta_\pm}}{\vert
m\vert\pi}\frac{n!k!}{\Gamma(n+L+\tfrac{d}{2})\Gamma(1+k\pm\nu)}\,.
\end{equation}

The sign function on the right hand side of
\eqref{creationannihilationcommutator} can be understood as follows.
The Fock space vacuum $\vert 0\rangle$ is defined by $a^\pm_{L,n,k}\vert
0\rangle=0$ for $m>0$ and ${a^\pm}^\dagger_{\!\!\!\!\!L,n,k}\vert
0\rangle=0$ for $m<0$. The interpretation of the latter statement is
that ${a^\pm}^\dagger_{\!\!\!\!\!L,n,k}$ for $m<0$ is the annihilation
operator for the antiparticle making $a^\pm_{L,n,k}$ for $m<0$ the
creation operator for the antiparticle. In lightcone quantisation it is common practise
to rename the creation and annihilation operators for $m<0$ by
$a^\pm_{-m,L,n,k}={b^\pm}^\dagger_{\!\!\!\!\!m,L,n,k}$ and likewise
${a^\pm}^\dagger_{\!\!\!\!\!-m,L,n,k}=b^\pm_{m,L,n,k}$ and restrict
$m$ to only take positive values. Here we will not use this notation
because $m$ is not summed over anyway. We could always restrict $m$ to
be positive; however, to test results we find it useful to keep track
of the sign of $m$. One other motivation for keeping both signs of $m$
comes from the fact that from these results 
one can obtain the results for scalar
field propagation on AdS in plane wave coordinates (after
setting $\beta=0$ and summing over all values of $m$).

For the normalisable modes $\phi_+$ the Hamiltonian $H_T$ is conserved
in time. For the normalisable modes $\phi_-$ this is not the case and
for these modes the action \eqref{eq:action} and the charges $K_A$
\eqref{eq:charges} are not appropriate. For the normalisable modes
$\phi_-$ following \cite{Breitenlohner:1982jf,Minces:2001zy} we expect
it to be necessary to introduce non-minimal coupling terms for the
scalar field $\phi$. This being said we stress that the condition
\eqref{eq:noflux1} is only a condition on the modes and is therefore
insensitive to the addition of non-minimal coupling terms in the bulk
and boundary action of \eqref{eq:action} since on-shell the Ricci scalar
is constant and can be absorbed into the definition of $m_0$.

\subsection{Time evolution}

We show in this section that even though the future domain of
dependence $\mathcal{D}^+(\Sigma_{T_0})$ of a constant $T$ slice
$\Sigma_{T_0}$ is empty, 
\be
\mathcal{D}^+(\Sigma_{T_0})=\emptyset
\ee
(section 3.1), the scalar field has a unique time
evolution that is fully predictable given appropriate initial data.

In the previous subsection we have identified two inequivalent Hilbert
spaces, those of the plus and minus modes $\phi^\pm$ 
\eqref{decomposition} respectively.
Both of the spaces satisfy the property
that any element is an eigenfunction of $N$. Let us denote these two
Hilbert spaces by $\mathcal{H}_m^+$ and $\mathcal{H}_m^-$.

We will show that for the Hilbert spaces with $m\neq 0$ there exists a
well-posed initial value problem in the sense that given initial data
for a scalar field in $\mathcal{H}_m^\pm$ at some time $T=T_0$ it is
possible to uniquely predict the future dependence. To see this one
just has to note that from $\phi(T=T_0,V,R, \vec X)$ and the mode
decomposition \eqref{decomposition} it is possible to read off the
coefficients $a_{L,n,k}$ via\footnote{We have used the following two
orthogonality relations: $\int
d\Omega_{d-1}Y_{L'}(\Omega)Y_L(\Omega)=\delta_{LL'}$ and
$\int_0^\infty dx x^a
e^{-x}L_n^a(x)L^a_{n'}(x)=\tfrac{\Gamma(n+a+1)}{n!}\delta_{nn'}$ where
$\text{Re}\,a>-1$.}
\begin{equation}
 \langle
e^{-imV}v^\pm_{L,n,k}\vert\phi(T=T_0)\rangle=\text{sign}(m)a^\pm_{L,n,k
}\,.
\end{equation}
Knowing all the $a^\pm_{L,n,k}$ determines the full future dependence
of the function $\phi$ (from \eqref{decomposition}). Note that in order to 
have a well-defined time 
evolution we only need to specify the values of the field $\phi$ at time 
$T=T_0$ and not its $T$-derivative.

This structure and property of the initial value problem and time-evolution
of scalar fields on Schr\"odinger space-times is preserved by the foliation-preserving
diffeomorphisms \eqref{fpd}. In any coordinate system obtained in this way, 
the Klein-Gordon equation is a 1st order differential equation in the new time coordiante
$T'$, and the evolution
of the Klein-Gordon field $\phi$ is determined by the value of the
field on the null surface $\Sigma_{T'}$ (and the momentum in the $V'$-direction, the mass).

We have thus resolved the problem associated with the emptiness of the
future domain of dependence $\mathcal{D}^+(\Sigma_{T_0})$. The emptiness of
$\mathcal{D}^+(\Sigma_{T_0})$ resulted from 1) the existence of lightlike
lines and 2) from the existence of curves that reach the boundary
before crossing the equal time surface. The way the scalars get around
this potential unpredictability follows from i) the restriction to
modes with $m\neq 0$ (as explained above, the $m=0$ modes are the scalar analogues of 
lightlike lines, and the restriction to $m\neq 0$ indeed avoids the problems associated with these 
lightlike lines), and ii) from imposing suitable boundary conditions which forbid
information exchange with the boundary.

\subsection{Wightman functions and Green's functions}

We will now first study the positive and negative frequency Wightman
functions, $G^\pm(x,x')$, and then from those build the bulk-to-bulk
propagator in global coordinates. We have
\begin{eqnarray}
 G^+(x,x') & = & \langle 0\vert\phi(x)\phi^\dagger(x')\vert
0\rangle\,,\\
  G^-(x,x') & = & \langle 0\vert\phi^\dagger(x')\phi(x)\vert 0\rangle\,.
\end{eqnarray}
Our conventions for the creation and annihilation operators are given
in \eqref{creationannihilationcommutator}. 
The positive and negative frequency Wightman functions, denoted by $G^+$
and $G^-$ respectively, can be defined for both Hilbert spaces
$\mathcal{H}_m^\pm$ where the $\pm$ refer to the two different sets of
normalisable modes in \eqref{decomposition}. We will write the expressions for $G^+$
and $G^-$ on $\mathcal{H}_m^\pm$ simultaneously, hoping that this does not
cause any confusion.
Using the mode
decompositions \eqref{decomposition} we obtain for the Wightman
functions the expressions
\begin{eqnarray}
 G^+(x,x') & = &
\tfrac{1}{2}\theta(m)e^{-im(V-V')}\sum_{L,n,k}(C^\pm_{L,n,k})^2e^{
-i2\omega\left(n+k+\tfrac{L}{2}+\tfrac{\Delta_\pm}{2}\right)(T-T')}
\times \nonumber\\
&& \times
Y_L(\Omega)Y^*_L(\Omega')\varphi_{L,n}(\rho)\varphi_{L,n}
(\rho')\phi_k^\pm(R)\phi_k^\pm(R')\,,\label{G+}\\
G^-(x,x') & = &
\tfrac{1}{2}\theta(-m)e^{-im(V-V')}\sum_{L,n,k}(C^\pm_{L,n,k})^2e^{
i2\omega\left(n+k+\tfrac{L}{2}+\tfrac{\Delta_\pm}{2}\right)(T-T')}
\times \nonumber\\
&& \times
Y_L(\Omega)Y^*_L(\Omega')\varphi_{L,n}(\rho)\varphi_{L,n}
(\rho')\phi_k^\pm(R)\phi_k^\pm(R')\,.\label{G-}
\end{eqnarray}
Both $G^\pm$ are solutions
to the homogeneous Klein--Gordon equation. We have under complex
conjugation $(G^\pm(x,x'))^*=G^\pm(x',x)$. As it stands the sums in
the expressions for $G^\pm$ are not convergent in the sense of
functions. If we consider the various sums as series in the parameter
$s=\exp[-i2\omega(T-T')]$ then the series only converges if $\vert
s\vert<1$. Thus, in order to have convergent series we replace $T-T'$ in
$G^+$ by $T-T'-i\epsilon$ and $T-T'$ in $G^-$ by
$T-T'+i\epsilon$,  with $\epsilon >0$ infinitesimal.
In terms of 
\begin{equation}
 s_\epsilon=e^{-i2\omega(T-T')-2\omega\epsilon}\,.
\end{equation}
the regulated $G^+$ is then a series in $s_\epsilon$ and the regulated $G^-$ is a series in $s_\epsilon^*$.

In order to evaluate the sums we use the following generating function
for the Laguerre polynomials (see e.g. \cite[Theorem 69]{rainville} or \cite{mos})
\begin{equation}
\sum_{n=0}^\infty
e^{-\tfrac{1}{2}(x+y)}\frac{(xy)^{\tfrac{a}{2}} s^n
n!}{\Gamma(n+a+1)}L_n^a(x)L_n^a(y)
=\frac{s^{-\tfrac{a}{2}}}{1-s}\exp[-\tfrac{1}{2}(x+y)\tfrac{1+s}{1-s}
]e^{-i\tfrac{\pi}{2}a}J_a(2i\tfrac{\sqrt{xys}}{1-s})\,.
\end{equation}
We will also need the decomposition of a plane wave into spherical
harmonics which is given by (see e.g. \cite{abas})
\begin{equation}
e^{iz\hat n\cdot\hat n'}=(2\pi)^{\tfrac{d}{2}}\sum_L i^L
z^{-\tfrac{d-2}{2}}J_{L+\tfrac{d-2}{2}}(z)Y^*_L(\hat n)Y_L(\hat n')\,,
\end{equation}
where $\hat n$ and $\hat n'$ are unit vectors on $S^{d-1}$ that are
parametrised by $\Omega$ and $\Omega'$, respectively. In fact, the
unit vector $\hat n$ is nothing but the Cartesian vector $\vec X$ that
appears in global Schr\"odinger metric normalised to unit length.

Armed with these two expressions we can evaluate the sums that define
the Wightman functions. The result is
\begin{eqnarray}
G^+(x,x') & = & \theta(m)\frac{i^{-\Delta_\pm}}{(2\pi)^{\tfrac{d}{2}}
4\pi m}(m\zeta_{-\epsilon})^{\tfrac{d+2}{2}}J_{\pm\nu}(m\zeta_{
-\epsilon} )e^{im\eta_{-\epsilon}}\,,\\
G^-(x,x') & = &
-\theta(-m)\frac{i^{\Delta_\pm}}{(2\pi)^{\tfrac{d}{2}}4\pi
m}(-m\zeta_{+\epsilon})^{\tfrac{d+2}{2}}J_{\pm\nu}(-m\zeta_{+\epsilon}
)e^{im\eta_{+\epsilon}}\,,\nonumber\\
&&
\end{eqnarray}
where $\zeta_{\pm\epsilon}$ and $\eta_{\pm\epsilon}$ are $\epsilon$-deformations 
of the invariant functions $\zeta(x,x')$ and $\eta(x,x')$ 
\eqref{etazeta} expressed in global coordinates. We have
\begin{eqnarray}
\zeta_{\pm\epsilon} & = & \frac{\omega RR'} {\sin\omega(T-T'\pm
i\epsilon)}\,,\label{Wightman+}\\
\eta_{\pm\epsilon} & = & -(V-V')+\frac{\omega(\vec X^2+\vec X'^2+
R^2+R'^2)}{2\tan\omega(T-T'\pm i\epsilon)}-\frac{\omega\vec X\cdot\vec
X'}{\sin\omega(T-T'\pm i\epsilon)}\,.\label{Wightman-}
\end{eqnarray}
It can be checked that, apart from the $i\epsilon$ and the overall constant, the result for the Wightman functions
agrees with the most general normalisable solution to the
Klein--Gordon equation for a function that only depends on
$\eta$ and $\zeta$. The Poincar\'e coordinate expressions for the
Wightman functions can be obtained by taking the $\omega\rightarrow 0$
limit in \eqref{Wightman+} and \eqref{Wightman-}.

Now that we have the two Wightman functions at our disposal we are in
a position to compute any Green's function that we are interested in.
For example the Feynman propagator is given by
\begin{equation}\label{propagator}
 G_F(x,x')=\theta(T-T')G^+(x,x')+\theta(T'-T)G^-(x,x')\,,
\end{equation}
and the retarded and advanced Green's functions read
\begin{eqnarray}
 G_R(x,x') & = & \theta(T-T')\left(G^+(x,x')-G^-(x,x')\right)\,,\\
 G_A(x,x') & = & \theta(T'-T)\left(G^+(x,x')-G^-(x,x')\right)\,,
\end{eqnarray}
where $G^+(x,x')-G^-(x,x')$ is called the commutator function. 

It is clear, though, that in the Schr\"odinger case, due to the fact
that $m$ is not summed over, there is no mixing between positive and
negative frequency Wightman functions. For example, for $m>0$ the
propagator and the retarded Green's functions are the same, while for
$m<0$ the propagator equals the advanced Green's function.

The fact that in the Feynman propagator the step function
$\theta(T-T')$ is multiplied by the step function
$\theta(m)$ appearing in the Wightman function $G^+$ and
similarly the fact that $\theta(T'-T)$ multiplies $\theta(-m)$
appearing in $G^-$ has the following welcome consequence. Even though
$T$ is not a global time function and as such does not allow one to
label all causally related events by a different value of $T$, it is
not a problem to define a time ordering since the time ordering in the
Feynman propagator is correlated with the sign of $m$. The failure of
$T$ to provide a well-defined global time ordering only applies to
events with the same value of $T$. Propagation between such events
with $m>0$ or $m<0$ does not occur.

The bulk-to-bulk propagator $G_F(x,x')$, \eqref{propagator}, satisfies
the delta-function sourced Klein--Gordon equation
\begin{equation}
 \left(\square-\tilde
m_0^2\right)G_F(x,x')=\frac{i}{2\pi}e^{-im(V-V')}R^{d+1}
\delta(T-T')\delta(R-R')\delta(\vec X-\vec X')\,.
\end{equation}
The bulk-to-bulk propagator for the Schr\"odinger space-time has also
been constructed in \cite{Volovich:2009yh}. Our result agrees with the
expression in \cite{Volovich:2009yh}.\footnote{However, 
equation (3.27) of \cite{Volovich:2009yh} contains a misprint. The
normalisation constant which they denote by $\tilde C_\Delta$ should
be the one given in \eqref{eq:AdSnormconst}. This latter normalisation
constant agrees with the one in \cite{D'Hoker:2002aw}.}

We next approximate the bulk-to-bulk propagator for points that are
close to being separated by a lightlike line, i.e. for $T-T'$ small,
and show how it is related to the Feynman propagator for a massless
particle on Minkowski space-time. Using the asymptotic form of the
Bessel function we find that for $T-T'$ small the bulk-to-bulk
propagator can be approximated by
\begin{eqnarray}
 G_F(x,x') & = &
\theta(m)\theta(T-T')\frac{1}{2}\frac{i^{-\tfrac{d+1}{2}}m^{\tfrac{d-1
}{2}}}{(2\pi)^{\tfrac{d+3}{2}}}\left(\frac{RR'}{T-T'-i\epsilon}
\right)^{\tfrac{d+1}{2}}e^{-im\alpha_-}\nonumber\\
&&+\theta(-m)\theta(T'-T)\frac{1}{2}\frac{i^{\tfrac{d+1}{2}}(-m)^{
\tfrac{d-1}{2}}}{(2\pi)^{\tfrac{d+3}{2}}}\left(\frac{RR'}{
T-T'+i\epsilon}\right)^{\tfrac{d+1}{2}}e^{im\alpha_+}\,,\nonumber\\
&&
\end{eqnarray}
where $\alpha_\pm$ is
\begin{equation}
 \alpha_\pm=\mp i\left(V-V'-\frac{1}{2}\frac{(\vec X-\vec
X')+(R-R')^2}{T-T'\pm i\epsilon}\right)\,.
\end{equation}
First of all notice that the expression is independent of $\beta$.
Secondly, the relation with the propagator for a massless particle on
Minkowski space-time is obtained by integrating this result over $m$.
Doing so we find
\begin{equation}\label{eq:Minkowskipropagator}
 \int_{-\infty}^\infty dm
G_F(x,x')=\frac{1}{\text{Vol}\,S^{d+2}}\frac{1}{d+1}
(\sigma+i\epsilon)^{-\tfrac{d+1}{2}}\,,
\end{equation}
where
\begin{equation}
 \text{Vol}\,S^{d+2}=\frac{2\pi^{\tfrac{d+3}{2}}}{\Gamma(\tfrac{d+3}{2
})}\,.
\end{equation}
In obtaining this expression we used that $\sigma$ is well
approximated by the Minkowski space-time geodesic distance for
lightlike separated points. Equation \eqref{eq:Minkowskipropagator}
is the standard expression for the propagator of a massless particle
on Minkowksi space-time. We thus conclude (by inverse Fourier
transform) that the behavior of the Schr\"odinger bulk-to-bulk
propagator for points that are close to being separated by a lightlike
line is well approximated by the Minkowski space-time propagator for a
massless particle with a fixed  momentum $m$ in the $V$ direction.

Information about the causal structure probed by scalars can be
obtained by looking at the zeros of the commutator function
$G^+(x,x')-G^-(x,x')$. By microcausality, the commutator function must
vanish for spacelike separated points $x$ and $x'$. In a free field
theory the commutator function is a classical $c$-number
quantity. Hence, it can only be nonzero whenever two points can be
connected by a classical path. The commutator function is therefore
sensitive to the possible geodesic non-connectedness. It follows
that the commutator function $G^+(x,x')-G^-(x,x')$ must be zero
when
\begin{enumerate}
 \item $x$ and $x'$ are spacelike separated (microcausality),
\item $x$ and $x'$ cannot be connected by a geodesic.
\end{enumerate}
Since $G^+$ only exists for positive values of $m$ and $G^-$ only for
negative values of $m$, the commutator function vanishes if 
and only if $G^\pm$ vanish separately. Below we will discuss these two
types of zeros of $G^\pm$.
For a recapitulation of the properties of the commutator function in
the AdS case which shows similar behaviour we refer to appendix
\ref{app:AdScommutatorfunction}.

Any two points for which $T-T'\neq 0$ are timelike separated. Hence
all spacelike separated points are points for which necessarily
$T=T'$ (section 3.2). In appendix \ref{app:geodesics} it is shown that points $P$
and $\bar P$ for which $T_{\bar P}-T_P=\tfrac{\pi}{\omega}$ that do not satisfy
\eqref{eq:nearlyantipodal} are geodesically disconnected. It
follows that, by points 1 and 2 above, the
commutator function must vanish whenever $\sin\omega(T-T')=0$. It can
be checked that the $i\epsilon$ prescription in the Wightman function
$G^\pm$ is precisely such that this is the case. 
Summarising we can say that the commutator function probes the
following part of the space-time
\begin{equation}
\label{sheet}
\bigcup_{n\in\ZZ} I^+(T=T'+(n-1)\tfrac{\pi}{\omega})\cap
I^-(T=T'+n\tfrac{\pi}{\omega}) 
\end{equation}
which, as we will now discuss, is the scalar field counterpart of the non-distinguishing
character of space-time as seen by point particle probes.

The boundary of the region on which the commutator function is
nonvanishing is given by $\sin\omega(T-T')=0$. To contrast this with the
AdS case note that there the commutator function is nonvanishing for
$\vert\eta^{\text{AdS}}\vert<1$ (see appendix E). In both cases the
boundaries are formed by lightlike lines. However, in AdS all lightlike
lines are null geodesics and these form a relativistic lightcone structure
whereas in the Schr\"odinger case only null geodesics with $P_-=0$ 
(see appendix \ref{app:geodesics}) form
lightlike lines, and these describe a Galilean lightcone structure. In the case of
massive point particles we saw that they probe the entire
chronological past and future $I^-(p_0)\cup I^+(p_0)$ of some point $p_0$ \eqref{p0}. The fact that the
propagator only probes the horizontal sheets \eqref{sheet} rather than
$I^-(p_0)\cup I^+(p_0)$ is something that is also observed in the case
of the propagator for the non-relativistic harmonic oscillator.

\subsection{Bulk-to-boundary propagator}

The bulk-to-boundary propagator $K_F$ can be obtained from:
\begin{equation}
 K_F(T,V,R,\vec X;T',V',\vec X')=C\lim_{R'\to
0}R'^{-\Delta_+}G_F(x,x')\,,
\end{equation}
where $G_F(x,x')$ is the bulk-to-bulk propagator depending on
$\Delta_+$. Using the expression for the bulk-to-bulk propagator we
find
\begin{eqnarray}
 K_F(T,V,R,\vec X;T',V',\vec X') & = & \nonumber\\
&&\hskip
-4cm\theta(m)\theta(T-T')C\frac{i^{-\Delta_+}m^{\Delta_+-1}}{4\pi
(2\pi)^{\tfrac{d}{2}}}\left(\frac{\omega
R}{\sin\omega(T-T'-i\epsilon)}\right)^{\Delta_+}
e^{im\eta_{-\epsilon}(R'=0)}\\
&&\hskip
-4cm+\theta(-m)\theta(T'-T)C\frac{i^{\Delta_+}(-m)^{\Delta_+-1}}{4\pi
(2\pi)^{\tfrac{d}{2}}}\left(\frac{\omega
R}{\sin\omega(T-T'+i\epsilon)}\right)^{\Delta_+}
e^{im\eta_{+\epsilon}(R'=0)}\nonumber\,.
\end{eqnarray}
The constant $C$ is determined by requiring (in the sense of
distributions):
\begin{equation}
 \lim_{\epsilon,R\to 0}R^{\Delta_+-d-2}K_F(T,V,R,\vec X;T',V',\vec
X')=\frac{1}{2\pi}e^{-im(V-V')}\delta(T-T')\delta(\vec X-\vec X')\,.
\end{equation}
In taking the limit we keep $\tfrac{T-T'}{R^2}$ and $\tfrac{\vec
X-\vec X'}{R}$ fixed as $R$ goes to zero and furthermore
$\tilde\epsilon\equiv\tfrac{\epsilon}{R^2}$ goes to zero as both
$\epsilon$ and $R$ go to zero. The result is that the constant $C$ is
given by
\begin{equation}
 C=\frac{i2^{1-\nu}}{\Gamma(\nu)}\,.
\end{equation}
The normalisation of the bulk-to-boundary propagator agrees with the
corresponding expression in \cite{Volovich:2009yh,Leigh:2009eb}. The
Poincar\'e coordinate expression for the bulk-to-boundary propagator
can be obtained by taking the $\omega\rightarrow 0$ limit.

When it comes to the bulk-to-boundary propagator there appears an
asymmetry in the discussion of the solutions depending on $\Delta_+$
and those in which $\Delta_+$ is replaced by $\Delta_-$. This also
happens in AdS and has to do with the fact the bulk-to-boundary
propagator with $\Delta_+$ replaced by $\Delta_-$ does not 
approach a boundary delta function  in the
limit where both points lie on the boundary.

The boundary value of the scalar field $\phi(T,V,R, \vec X)$ will be
denoted by $\phi_0(T,V,\vec X)$ and is defined by
\begin{equation}
 \phi_0(T,V,\vec X)=\lim_{R\to 0}R^{\Delta_+-d-2}\phi(T,V,R,\vec X)\,.
\end{equation}
A solution to the Klein-Gordon equation for a massive
complex scalar on the Schr\"odinger space-time for a normalisable mode
in the background of a non-normalisable mode is given by
\begin{eqnarray}
 \phi(T,V,R, \vec X) & = & \int dT'd^d\vec X'dV'K_{F}(T,V,R,\vec
X;T',V',\vec X')\phi_0(T',V',\vec X')\nonumber\\
&&+\phi_+(T,V,\vec X,R)\,,
\end{eqnarray}
where $\phi_+(T,V,\vec X,R)$ is given in \eqref{decomposition}. The
solution $\phi_+(T,V,\vec X,R)$ corresponds
to the normalisable solution \eqref{decomposition} while the 
part involving the bulk-to-boundary propagator corresponds to
the non-normalisable solution ($\phi_0$ is the boundary value of a
non-normalisable solution). The non-normalisable solution contains
both terms proportional to $R^{\Delta_-}$ as well as terms
proportional to $R^{\Delta_+}$ in the near boundary expansion of the
scalar field. The normalisable solution only contributes to the term
$\propto R^{\Delta_+}$.

When $\nu>1$ the term $\propto 
R^{\Delta_-}$ in the near boundary expansion of the scalar field is dual to a source in the boundary theory.
As is well-known when $0<\nu<1$ it is possible to instead consider the term 
$\propto R^{\Delta_+}$
as dual to a source. In the AdS/CFT context
the terms proportional to $R^{\Delta_+}$ and
$R^{\Delta_-}$ are conjugate variables in the sense that the
generating functional for the theory in which the term 
$\propto R^{\Delta_+}$ acts as the source can be obtained from the theory in
which the on-shell action depends on the term
$\propto R^{\Delta_-}$ via a Legendre transformation \cite{Klebanov:1999tb} (see also \cite{Marolf:2006nd} for the case of Lorentzian AdS/CFT).
We expect that a suitably modified version of this statement applies
here as well, so that it is sufficiently general to consider only the
case where the term proportional to $R^{\Delta_-}$ is dual to the
source and hence resides in the non-normalisable solution.

\section{Discussion}

We studied in detail the causal structure of the $z=2$ Schr\"odinger
space-time from the point of view of both point particle and scalar field
probes, emphasising and highlighting those peculiar features of the point particle
causal structure that have a counterpart for scalar fields.
For scalar fields, it turns out that the restriction to a fixed non-zero lightcone momentum $m$
(as dictated by the representation theory of the Schr\"odinger group) 
is sufficient to avoid the occurence of near-to-pathological
properties that one does encounter in the case of point particle
probes. For example, even though one cannot define a time function
and even though the future domain of dependence of slices $\Sigma_T$ of constant
global coordinate time $T$ are empty, one can define a well-posed initial
value problem for scalar fields. We have shown that, for a given $m$ this requires
specification of the scalar field on $\Sigma_T$.
This first-order nature of the Klein-Gordon initial value problem is preserved by the
so-called double-foliation preserving diffeomorphisms which leave the Galilean-like
causal structure (and the lightlike lines) of the $z=2$ Schr\"odinger space-time invariant.
This Galilean-like causal structure, as defined by the properties of
causal curves, is reflected in
the properties of the Wightman functions and
propagators of the scalar field theory.

One obvious extension of this work is to consider Schr\"odinger
space-times with values of $z$ different from two. The range of $z$
that is interesting from the point of view of non-relativistic
physics is $z>1$. From the study of tidal forces we know that
for $1<z<2$ the space-times are singular \cite{bhr}. This leaves us with
the range $z>2$. In this case one would like to construct the counterpart
of the $z=2$ global metric. The construction of such a global metric
is hampered by the non-existence of an everywhere timelike Killing
vector which means that any global coordinate system is necessarily
time-dependent. In order to find an explicit global metric one could
try to generalize the isometric embedding presented here to other
values of $z$. This is indeed possible but the result for us was not
sufficiently illuminating to derive from it a global metric. What can
be stated just from knowing the Poincar\'e like coordinates for $z>2$
is that these space-times are non-distinguishing. This can be proven
using an appropriately adapted verion of the curve given in \cite{hubeny}.

It would also be interesting to study metric perturbations. This
is relevant for a number of reasons. First of all, we know that since
the $z=2$ Schr\"odinger space-time is not stably causal there exist
perturbations of the lightcone structure that lead to the existence of
closed timelike curves. This raises a number of questions: what kind of
metric perturbations produce this kind of behavior? are these physically
relevant (e.g.\ do the perturbations have finite energy in a suitable
sense)? and how sensitive would scalar fields be to the presence of
such closed timelike curves in the perturbed metric?  Secondly, in the
analysis of the scalars, representation theory played a dominant role
(choosing a fixed nonzero $m$). It would be nice to understand what
this entails for the metric perturbations. Ultimately one would like to
understand the precise form of the asymptotic fall-off conditions for the
various fields (scalar, gauge, metric, etc.) and the required counterterms
that allow one to define
a well-defined variational problem and understand the construction of
holographic renormalisation (see \cite{hrr,Compere:2009qm,ross} for a
discussion of some of the issues involved). These issues are also 
relevant for the study of asymptotically Schr\"odinger black holes. 
In particular, one might like to understand whether or not black holes 
in global Schr\"odinger exhibit any interesting phase transitions \cite{hir}.

\subsection*{Acknowledgements}

This work has been supported by the Swiss National Science Foundation and
the ``Innovations- und Kooperationsprojekt C-13'' of the Schweizerische
Universit\"atskonferenz SUK/CUS. 

\newpage

\appendix


\section{Schr\"odinger algebra and isometric embeddings}\label{sec:Schalgebra}
\label{app:nogotheorem}

We will denote the isometry algebra of the Schr\"odinger 
space-time $\mathsf{Sch}_{d+3}$ by $\mathfrak{sch}(d)$.
It consists of all the elements of the isometry algebra
$\mathfrak{so}(2,d+2)$ of $\mathsf{AdS}_{d+3}$ that commute with
the lightcone momentum $P_-$: $d$-dimensional
spatial rotations $M_{ab}$ and translations $P_a$, Galilean
boosts $V_a$, time translations $H$, dilatations $D$, a special
conformal transformation $C$ and, of course, the central element 
$P_-\equiv N$. The latter four generators $\{H,C,D,N\}$
form the algebra
\be
\begin{aligned}
&\mathfrak{sch}(d=0)\cong \mathfrak{sl}(2,\RR)\oplus \RR_N
\cong \mathfrak{so}(2,1)\oplus \RR_N\\
&{}[H,C] = D\qquad [D,C]=2C \qquad [D,H] = -2H\;\;.
\end{aligned}
\ee
The other non-trivial commutators are
\be
\begin{aligned}
&\left[D,P_a\right] = -P_a \quad \left[D,V_a\right] = V_a \quad 
\left[P_a,V_b\right] = \delta_{ab}N \quad
\left[H,V_a\right] = P_a  \quad \left[C,P_a\right] = -V_a \\
&
\left[M_{ab},P_c\right] = \delta_{bc}P_a- \delta_{ac}P_b\qquad 
\left[M_{ab},V_c\right] = \delta_{bc}V_a- \delta_{ac}V_b  \\
&
\left[M_{ab},M_{cd}\right] = \delta_{bc}M_{ad}+\delta_{ad}M_{bc}
-\delta_{bd}M_{ac}-\delta_{ac}M_{bd}
\end{aligned}
\ee
In Poincar\'e coordinates a realisation of this algebra is given by
\be
\label{akv}
\begin{aligned}
&H=\del_t \qquad P_a = \del_a \qquad 
V_a = x^a\partial_\xi + t\partial_{a} \qquad
M_{ab} = x^a\partial_{b} - x^b\partial_{a}\\
&N=\del_\xi \qquad D = 2t\partial_t + r\partial_r + x^a \del_a\qquad
C = t^2\partial_t
+\trac{1}{2}(r^2+\vec{x}^2)\partial_\xi+tr\partial_r+tx^a\partial_{a} \\
\end{aligned}
\ee
The embedding of $\mathfrak{sch}(d)$ into the AdS isometry algebra
$\mathfrak{so}(2,d+2)$ proceeds principally via the essentially
unique embedding of 
$\mathfrak{sch}(d=0)$ into $\mathfrak{so}(2,2)$ via the double null
splitting 
\be
\label{2121}
\mathfrak{so}(2,1)\oplus \RR_N \hookrightarrow 
\mathfrak{so}(2,1) \oplus \mathfrak{so}(2,1) \cong \mathfrak{so}(2,2)\;\;.
\ee
Explicitly, in terms of the generators
$M_{AB}$ of $\mathfrak{so}(2,d+2)$ satisfying
\be
\begin{aligned}
&\left[M_{AB},M_{CD}\right] = \eta_{BC}M_{AD}+\eta_{AD}M_{BC}
-\eta_{BD}M_{AC}-\eta_{AC}M_{BD}\\
&\eta_{AB}=\text{diag}(-1, +1, \ldots,+1, -1)\qquad A,B = 0,1, \ldots, d+3
\end{aligned} 
\ee
and null coordinates $(x^0,x^1)\ra x^\pm$, $(x^{d+2},x^{d+3}) \ra x^{\hat{\pm}}$ with 
$\eta_{+-}=\eta_{\hat{+}\hat{-}}=1$, one can choose
\be
D= M_{+-}+ M_{\hat{+}\hat{-}} \qquad H = M_{-\hat{-}}\qquad C = M_{+\hat{+}}
\ee
and $N$ any element of the other (commuting) $\mathfrak{so}(2,1)\subset\mathfrak{so}(2,2)$ \eqref{2121}.

In particular, if one seeks a codimension 1 embedding of $\mathsf{Sch}_{3}$
into $\RR^{2,2}$ (i.e.\ the Schr\"odinger analogue of the standard embedding
$\mathsf{AdS}_3 \hookrightarrow \RR^{2,2}$), the equation defining the hypersurface
has to be invariant under $\{H,C,D,N\}$ thought of as elements of $\mathfrak{so}(2,2)$
via the above embedding of Lie algebras. Thus let $f=f(x^\pm, x^{\hat\pm})$ be such a function.
Invariance under $H$, say, requires $f$ to satisfy the equation
\be
\left(x^+ \del_{\hat{-}}-x^{\hat{+}}\del_-\right)f(x^\pm,x^{\hat\pm})=0\;\;,
\ee
which is solved by $f=f(x^+,x^{\hat{+}},x^+x^- + x^{\hat{+}}x^{\hat{-}})$. Likewise, invariance under
$C$ relates the $x^+$- and $x^{\hat{+}}$-dependence, and one immediately finds
\be
Hf = Cf = 0 \quad\Ra\quad f=f(x^+x^- + x^{\hat{+}}x^{\hat{-}}) \;\;.
\ee
But since 
\be
x^+x^- + x^{\hat{+}}x^{\hat{-}}= -(x^0)^2 + (x^1)^2 + (x^{d+2})^2 - (x^{d+3})^2 
\ee
this hypersurface describes $\mathsf{AdS}_{3}$ with the enhanced isometry algebra $\mathfrak{so}(2,2)
\supsetneq \mathfrak{sch}(d=0)$. This argument immediately carries over to $d>0$ to preclude the existence
of an isometric embedding $\mathsf{Sch}_{d+3}\hookrightarrow\RR^{2,d+2}$. 

We are thus lead to consider codimension 2 embeddings $\mathsf{Sch}_{d+3}\hookrightarrow\RR^{2,d+3}$, 
with $\eta_{d+4,d+4}=1$, and the corresponding embedding of isometry algebras
\be
\mathfrak{sch}(d) \hookrightarrow \mathfrak{so}(2,d+3)\oplus \RR^{2,d+3}\;\;, 
\ee
in particular $\mathfrak{so}(2,1)\oplus \RR_N \hookrightarrow \mathfrak{so}(2,3)\oplus \RR^{2,3}$. A
characteristic feature of the Schr\"odinger algebra and Schr\"odinger geometry is the existence of the
central element $N$ realised as a null Killing vector. Thus $N$ can either arise from a null translation
in the translational part of the isometry algebra of the embedding space or from a null rotation. Let us
first show that the former is not possible (and that in fact the entire Schr\"odinger algebra needs to be
embedded into the rotational part  $\mathfrak{so}(2,d+3)$ of the isometry algebra). The argument is largely
insensitive to the dimension and signature of the embedding space, so we can consider a general (semi-direct
product) isometry algebra $\mathfrak{so}(p,q)\oplus \RR^{p,q}$ with $p+q$ large enough to accommodate
the required translations, and we assume that $N=P_-$ is identified with a
null translation. Then 
\begin{itemize}
\item[*] to reproduce $[P_a,V_b]=\d_{ab}N$, $P_a$ and $V_b$ cannot both be simultaneously translations or
rotations, so we choose $P_a \in \RR^{p,q}$, $V_a \in \mathfrak{so}(p,q)$ (the opposite choice is related 
to this via the authomorphism $P_a \lra V_a, H\lra -C$, $D\ra -D$, $N\ra -N$);
\item[*] since $C$ and $D$ do not commute with $P_a$, they are elements of $\mathfrak{so}(p,q)$;
\item[*] since $[H,C]=D$, one also has $H \in \mathfrak{so}(p,q)$;
\item[*] but then $[H,V_a] \in \mathfrak{so}(p,q)$, which contradicts the relation $[H,V_a]=P_a$. 
\end{itemize}
Thus we need to choose $N\in \mathfrak{so}(p,q)$. But then the Schr\"odinger algebra requires all
generators to be elements of $\mathfrak{so}(p,q)$ (by similar reasoning), and we need to consider
the embedding of isometry algebras
\be
\mathfrak{sch}(d) \hookrightarrow \mathfrak{so}(2,d+3)\;\;, 
\ee
in particular $\mathfrak{so}(2,1)\oplus \RR_N \hookrightarrow \mathfrak{so}(2,3)$. 
In addition to the embedding
via $\mathfrak{so}(2,2) \subset \mathfrak{so}(2,3)$ discussed (and dismissed) above, there are the two regular
embedddings
\be
\begin{aligned}
&\mathfrak{so}(2,1)\oplus \RR_N \cong \mathfrak{so}(1,2) \oplus \mathfrak{so}(1,1) \subset \mathfrak{so}(2,3)\\
&\mathfrak{so}(2,1)\oplus \RR_N \cong \mathfrak{so}(2,1) \oplus \mathfrak{so}(2) \subset \mathfrak{so}(2,3)
\end{aligned}
\ee
However, in these cases $N$ is identified either with a timelike boost generator or a spacelike rotation
generator and can therefore not possibly be null in the metric induced from the metric on $\RR^{2,3}$
(the only embedding that allows a null $N$ is that via $\mathfrak{so}(2,2)$). This argument generalises
in an obvious way to $d>0$ (by first embedding the rotations into $\mathfrak{so}(d)\subset \mathfrak{so}(2,d+3)$
and then dealing with the commuting $\mathfrak{so}(2,1)\oplus \RR_N$ algebra as above).

We therefore conclude that there are no codimension 2 \textit{equivariant}
isometric embeddings of the Schr\"odinger space-time, i.e.\ isometric
embeddings which are such that all isometries are induced by the
linear isometries (pseudo-orthogonal transformations) of the flat
embedding space. In this context it is worth noting that there exist
$G$-equivariant versions of the Nash embedding theorem (such as the
Moore-Schlafly theorem \cite{ms}), but that these do not produce useful
upper bounds on the required dimension of the embedding space.

\section{Schr\"odinger invariants}

In this appendix we briefly discuss the Schr\"odinger analogue of
the AdS chordal distance (or any AdS invariant measure of the
distance of two points like the geodesic distance).  A characteristic
feature of the Schr\"odinger space-time is that, due to its reduced
isometry (and isotropy) algebra, there are two independent invariant
building blocks instead of just the one unique chordal distance in
the AdS case.  To see this,  let $\sigma(x,x')$ be any function of
two points $x$ and $x'$ which is invariant under the simultaneous
action of the isometry group on $x$ and $x'$,
\be
\sigma(gx,gx')=\sigma(x,x')
\ee
(geodesic distance is an example of such a function).  If we consider
$\sigma(x,x')$ as a function of $x$ only, keeping $x'$ fixed,
$f_{x'}(x) = \sigma(x,x')$, then this function is invariant under
the stabiliser $H_{x'}$ of the point $x'$.

Concretely in the case of the Schr\"odinger space-time \eqref{metrics}
with its Schr\"odinger isometry group, let us e.g.\ consider the
point $x'=(t',\xi',r',\vec x')=(0,\xi',1,\vec 0)$ with $\xi'$
arbitrary. Its stabiliser is generated by the Killing vectors that
vanish at that point. It is easily seen from \eqref{akv} that these
are the linear combinations of $\{C-\tfrac{1}{2}N,V_a,M_{ab}\}$
\cite{SchaferNameki:2009xr}, forming an algebra isomorphic to
$\mathfrak{euc}(d)\oplus \RR$ (with $\mathfrak{euc}(d)$ the Euclidean
algebra).  The most general function invariant under these Killing
vectors depends on two variables.  Indeed, starting with a function
of all $d+3$ coordinates,  rotation invariance reduces the number
to 4 (3+ radial coordinate in the $x^a$-directions).  Then boost
invariance reduces this further by one (by correlating the
$t$-dependence with the dependence on this radial coordinate) and
finally invariance under $C-\tfrac{1}{2}N$ reduces this to two.

Since the space-time is homogeneous, this counting argument gives
the same number at each point of the space-time. Hence, for each
point $x'$ the function $f_{x'}(x)$ depends on two variables.
Therefore any Schr\"odinger invariant function $\sigma(x,x')$ of
$x$ and $x'$ is parametrised by two Schr\"odinger invariant functions
that we denote as $\eta(x,x')$ and $\zeta(x,x')$.  In Poincar\'e
coordinates they can be choosen to be
\be
\label{etazeta}
\eta(x,x')  =  -(\xi-\xi')+\frac{r^2+r'^2+(\vec x-\vec x')^2}{2(t-t')}\quad,\quad
\zeta(x,x') =  \frac{rr'}{t-t'}\;\;.
\ee
In particular, the standard AdS-invariant chordal distance is
\be
\eta^{\text{AdS}} = \frac{\eta}{\zeta} = \frac{-2(\xi-\xi')(t-t') + r^2 + r'^{2} + (\vec{x}-\vec{x}')^2}{2rr'}
\;\;.
\ee

\section{Chronological future}\label{sec:chronologicalfuture}

Here we prove that the chronological future of an arbitrary point $p_0$ on the 
constant global time slice $\Sigma_{T_0}$ consists of all points in the
space-time with $T>T_0$. To do so we will show that any two
points $(T_0,V_0,R_0,\vec X_0)$ and $(T_0+\varepsilon,V_f,R_f,\vec X_f)$
with $\varepsilon$ arbitrary can be connected by a timelike curve. These
curves can be constructed in strict analogy to the curves that were
used in \cite{hubeny} to prove the non-distinguishing character of the
$z=3$ Schr\"odinger space-time in Poincar\'e coordinates. First we 
adapt the curves to the $z=2$ case, then we
simply replace the Poincar\'e coordinates
$(t,\xi,r,\vec x)$ by the global coordinates $(T,V,R,\vec X)$.
This produces a new curve which is not equivalent to the one used in Poincar\'e
coordinates by a coordinate transformation. Nevertheless, by construction, 
the new curve has the same nice properties as 
the one used in \cite{hubeny}: for any two points $P_0$ and $P_f$ with
$T_0\neq T_f$ (but possibly $T_f - T_0= \epsilon >0$ infinitesimal),
there exists a causal curve connecting these points.\footnote{Since 
$P_0$ and $P_f$ can be spatially arbitrarily close to each other,
there exist causal curves that get arbitrarily close to being closed
causal curves. This is a violation of strong causality.} Moreoever,  
as a consequence of the strictly positive terms proportional to $\omega^2$ 
appearing in the global metric, the curve produced in this way is now 
actually everywhere timelike (and not just causal).

For notational simplicity we give the curve in terms of its tangent 
and its intermediate points:
\begin{equation}
\gamma(\lambda) = \left\{
\begin{array}{llll}
\gamma_1(\lambda) & & \mbox{for} & \lambda \in
[0,\frac{\varepsilon}{4}] \\
\gamma_2(\lambda) & & \mbox{for} & \lambda \in
[\frac{\varepsilon}{4},\frac{\varepsilon}{2}] \\
\gamma_3(\lambda) & & \mbox{for} & \lambda \in
[\frac{\varepsilon}{2},\frac{3\varepsilon}{4}] \\
\gamma_4(\lambda) & & \mbox{for} & \lambda \in
[\frac{3\varepsilon}{4},\varepsilon]
\end{array}\right. \quad
\begin{array}{rclcl}
P_0 & = & (T_0,V_0,R_0,\vec{X}_0) & = & \gamma(0) \\
P_1 & = & (T_0+\frac{\varepsilon}{4},V_1,R_1,\vec{X}_0)& = &
\gamma(\frac{\varepsilon}{4}) \\
P_2 & = & (T_0+\frac{\varepsilon}{2},V_1,R_1,\vec{X}_f) & = &
\gamma(\frac{\varepsilon}{2}) \\
P_3 & = & (T_0+\frac{3\varepsilon}{4},V_2,R_1,\vec{X}_f)& = &
\gamma(\frac{3\varepsilon}{4}) \\
P_f & = & (T_0+\varepsilon,V_f,R_f,\vec{X}_f) & = &
\gamma(\varepsilon)\,
\end{array}
\end{equation}
\begin{equation}
\dot{\gamma}(\lambda) = \left\{
\begin{array}{llll}
\dot{\gamma}_1(\lambda) = \left(1,\frac{8(R_1-R_0)^2}{\varepsilon^2}-
\frac{\beta^2}{2R(\lambda)^2},\frac{4(R_1-R_0)}{\varepsilon},0\right) & & \mbox{for} & \lambda \in
[0,\frac{\varepsilon}{4}] \\
\dot{\gamma}_2(\lambda) = \left(1,0,0,\frac{4(\vec{X}_f-\vec{X}_0)}{\varepsilon}\right) 
& & \mbox{for} & \lambda \in
[\frac{\varepsilon}{4},\frac{\varepsilon}{2}] \\
\dot{\gamma}_3(\lambda) = \left(1,\frac{4(V_2-V_1)}{\varepsilon},0,0\right) & & \mbox{for} & \lambda \in
[\frac{\varepsilon}{2},\frac{3\varepsilon}{4}] \\
\dot{\gamma}_4(\lambda) = \left(1,\frac{8(R_f-R_1)^2}{\varepsilon^2}-
\frac{\beta^2}{2R(\lambda)^2},\tfrac{4(R_f-R_1)}{\varepsilon},0\right) & & \mbox{for} & \lambda \in
[\frac{3\varepsilon}{4},\varepsilon]
\end{array}\right. \qquad \quad \,
\end{equation}
One sees that $\gamma_1$ and $\gamma_4$ are timelike by 
construction without requiring anything else, while in
order for the curve to be timelike along the segments 
$\gamma_2$ and $\gamma_3$ one needs to satisfy 
the inequality in \eqref{causalconstraint}, leading to 
the conditions
\begin{equation}\label{condition} 
\begin{array}{rclcl}
\frac{\beta^2}{R_1^2}  + \omega^2\left(\vec{X}(\lambda)^2+
R_1^2\right)  & > & \frac{16(\vec{X}_f-\vec{X}_0)^2}{\varepsilon^2} & & \text{along}\; \gamma_2 \\
\frac{\beta^2}{R_1^2} + \omega^2\left(R^2_1+\vec{X}_f^2\right) & > &
\tfrac{8}{\varepsilon}(V_1-V_2) & & \text{along}\; \gamma_3 
\end{array}
\end{equation}
where $V_1-V_2$ can be expressed in terms of the arbitrary starting and end points as 
\begin{equation}
 V_1-V_2=V_0-V_f+\tfrac{2}{\varepsilon}(R_1-R_0)^2+\tfrac{2}{
\varepsilon}(R_f-R_1)^2-\beta^2\tfrac{\varepsilon}{8R_0
R_1}-\beta^2\tfrac{\varepsilon}{8R_1 R_f}\,.
\end{equation}
When $\beta\neq 0$, the conditions \eqref{condition}
can be satisfied for any beginning and endpoints $P_0$ and $P_f$ of
the curve, in particular for any given $\epsilon = T_f-T_0 \neq 0$, 
by choosing $R_1$ small enough (i.e.\ by taking the path connecting
the two points to go sufficiently close to the boundary at $R=0$).
We thus find
that the chronological future (past) of any point $(T_0,V_0,R_0,\vec
X_0)$ is the entire set of points  with $T>T_0$ ($T<T_0$). In particular, 
all points on an equal time slice $\Sigma_{T_0}$ have identical future
and past, and in this sense the space-time is \textit{maximally non-distinguishing}.
This argument also shows precisely how the construction of this curve, and hence the argument, 
breaks down for $\beta = 0$ (plane wave AdS).

\section{Geodesics}\label{app:geodesics}

In this appendix we describe those properties of the solutions to the
geodesic equations that are relevant for our purposes. We do not give
the explicit solutions to the geodesic equations. 

The geodesic equations are
\begin{eqnarray}
	\dot T & = & P_-R^2\,,\label{eq:Tgeodesicequation}\\
	\dot V & = & ER^2-\beta^2 P_--\omega^2P_-R^2(R^2+\vec
X{}^2)\,,\label{eq:Vgeodesicequation}\\
	\frac{1}{R^2}\frac{d}{d\lambda}\left(\frac{1}{R^2}\dot{\vec
X}\right) & = & -\omega^2P_-^2\vec
X\,,\label{eq:vecXgeodesicequation}\\
	k & = & \beta^2 P_-^2+\omega^2P_-^2R^4+(\vec
P{}^2-2P_-E)R^2+\frac{\dot
R^2}{R^2}\,,\label{eq:radialgeodesicequation}
\end{eqnarray}
where $E$, $P_-$ and $\vec P$ are integration constants and the dot
indicates differentiation with respect to $\lambda$ which depending on
$k=0,\pm 1$ is either proper time, proper length or some affine
parameter. $P_-$ is the lightcone momentum conjugate to $V$, 
and solutions to the geodesic equation with $P_-=0$ either have $k=0$
(these are lightlike lines - see section 3) or $k=1$. In this appendix we will always
assume that $P_-\neq 0$. There are three families of solutions that
depend on whether $\kappa=k-\beta^2 P_-^2$ is negative, zero or
positive. 

When $\beta=0$ we have $\kappa=k$ and the three cases
split into timelike, null and spacelike geodesics. When $\beta\neq 0$
this does not happen. Both the timelike and the null geodesics are
sitting in the $\kappa<0$ class of solutions, while the spacelike
geodesics are divided among all three classes with the $\kappa=0$ and
$\kappa>0$ classes containing only spacelike geodesics.

Geodesics with $\kappa<0$ describe bounded motion on $0<R<\infty$
and never reach the points $R=0$ and $R=\infty$. Since the $\kappa<0$
class of solutions also contains spacelike geodesics not all spacelike
geodesics go to the boundary. The motion for $\kappa<0$ is periodic
in the $R$ and $\vec X$ directions with periods $\pi/\omega$ and
$2\pi/\omega$, respectively. The motion in the $V$ direction
(for non-compact $V$) is however not periodic. This is due to the
second term (containing $\beta$) on the right hand side of
\eqref{eq:Vgeodesicequation}. This term would not be there in plane
wave AdS. For compact $V$ the periodicity of the motion in the $R$ and
$\vec X$ directions does not generically coincide with the periodicity
of identifications $V\sim V+2\pi L$.

All geodesics with $\kappa<0$ that go through some point $P$, say,
also go through the point $\bar P$ with coordinates
\be
\left(T_{\bar P},  V_{\bar P}, R_{\bar P}, X_{\bar P}^a\right)=
\left(T_P+\frac{\pi}{\omega}, V_P-\beta^2\Delta V, R_P, -X^a_P\right)\;\;, 
\label{eq:nearlyantipodal}
\ee
where $\Delta V$ is some $\beta$ independent difference that depends
on the locations of $P$ and $\bar{P}$ as well as on the parameters of the
geodesic connecting $P$ and $\bar{P}$. In AdS points $P$ and
$\bar P$ are examples of antipodal points.

It follows from the periodicity of the $\kappa<0$ class of geodesics
that points $P$ and $Q$ with $T_Q-T_P=\pi/\omega$ and with $R_Q\neq
R_P$ can never be connected by a $\kappa<0$ geodesic. Such points $P$
and $Q$ can also not be connected by $\kappa=0$ or $\kappa>0$
geodesics because those reach the boundary within a time interval of
$\pi/\omega$ or less. This proves that the Schr\"odinger
space-time (just as AdS) is not geodesically connected. 
In the case of AdS, the geodesic disconnectedness can be compactly
described in terms of the invariant distance $\eta^{\text{AdS}}$:
if $\eta^{\text{AdS}}(x,x')\le-1$ and $x'\neq\bar x$, then there 
is no geodesic connecting $x$ and $x'$. In particular, 
for $\beta = 0$ the above pair of points $P,Q$ provides an
example of such a pair of points 
since $\eta^{\text{AdS}}(x_P,x_Q)<-1$.

\section{AdS commutator function}\label{app:AdScommutatorfunction}

The AdS Wightman functions, $G^\pm_{\text{AdS}}(x,x')$, can be
obtained via
\begin{equation}
 G^\pm_{\text{AdS}}(x,x')=\int_{-\infty}^\infty dm
G^\pm_{\beta=0}(x,x')\,.
\end{equation}
In order to perform the integral over $m$ we allude to the following
result taken from \cite{Gradshteyn}
\begin{equation}
 \int_0^\infty dx e^{-\alpha x}J_\gamma(\beta
x)x^{\mu-1}=\frac{\left(\tfrac{\beta}{2\alpha}\right)^\gamma
\Gamma(\gamma+\mu)}{\alpha^\mu\Gamma(\gamma+1)}F\left(\tfrac{
\gamma+\mu}{2},\tfrac{\gamma+\mu+1}{2};\gamma+1;-\tfrac{\beta^2}{
\alpha^2}\right)\,,
\end{equation}
where $F$ is the hypergeometric function and where we must have
\begin{equation}\label{eq:conditions}
 \text{Re}\,(\mu+\gamma)>0\hskip 1cm\text{and}\hskip 1cm
\text{Re}(\alpha\pm i\beta)>0\,.
\end{equation}
By taking $x=m$, $\alpha=-i\eta_{-\epsilon}$,
$\beta=\zeta_{-\epsilon}$, $\gamma=\pm\nu$, and $\mu=\tfrac{d+2}{2}$
we obtain for $G^+_{\text{AdS}}$
\begin{equation}
 G^+_{\text{AdS}}(x,x')=C_{\Delta_\pm}\left(\eta^{\text{AdS}}_{
-\epsilon}\right)^{-\Delta_\pm}F\left(\tfrac{\Delta_\pm}{2},\tfrac{
\Delta_\pm+1}{2};\Delta_\pm-\tfrac{d}{2};(\eta^{\text{AdS}}_{-\epsilon
})^{-2}\right)\,,
\end{equation}
where $C_{\Delta_\pm}$ is given by
\begin{equation}\label{eq:AdSnormconst}
 C_{\Delta_\pm}=\frac{\Gamma(\Delta_\pm)}{2^{\Delta_\pm}\pi^{\tfrac{
d+2}{2}}(2\Delta_\pm-d-2)\Gamma(\Delta_\pm-\tfrac{d}{2}-1)}\,,
\end{equation}
and where $\eta^{\text{AdS}}_{-\epsilon}$ is given by
\begin{equation}
\eta^{\text{AdS}}_{-\epsilon}=\frac{\eta_{-\epsilon}}{\zeta_{-\epsilon
}}\,.
\end{equation}
Similarly, with $\alpha=i\eta_{+\epsilon}$ and the same choices for
$\beta$, $\gamma$ and $\mu$, we obtain for $G^-_{\text{AdS}}(x,x')$
the same expression as we have for $G^+_{\text{AdS}}$ but this time as
a function of
$\eta^{\text{AdS}}_{+\epsilon}=\frac{\eta_{+\epsilon}}{\zeta_{
+\epsilon}}$. Note that the $i\epsilon$ prescription is such that the
conditions \eqref{eq:conditions} for $\alpha$ and $\beta$ are
fulfilled. To see this more explicitly use the fact that to first
order in $\epsilon$ we have
\begin{eqnarray}
 \zeta_{\pm\epsilon} & = & \zeta\mp
i\omega^2\epsilon\frac{RR'\cos\omega(T-T')}{\sin^2\omega(T-T')}\,,\\
 \eta_{\pm\epsilon} & = & \eta\mp
i\frac{\omega^2\epsilon}{2}\frac{R^2+R'^2+\vec X^2+\vec X'^2-2\vec
X\cdot\vec X'\cos\omega(T-T')}{\sin^2\omega(T-T')}\,.
\end{eqnarray}

Consider the commutator function
$[\phi(x),\phi(x')]=G^+_{\text{AdS}}(x,x')-G^-_{\text{AdS}}(x,x')$. As
long as $\vert\eta_{\pm\epsilon}^{\text{AdS}}\vert>1$, where
$\eta_{\pm\epsilon}^{\text{AdS}}=\eta_{\pm\epsilon}/\zeta_{\pm\epsilon
}$, the hypergeometric function in $G^\mp_{\text{AdS}}$ is defined by
its series expansion and is thus single-valued for any $\epsilon$.
Since the series converges absolutely for
$\vert\eta^{\text{AdS}}\vert>1$, i.e. for $\epsilon=0$, we can take
the limit $\epsilon\rightarrow 0$ and we get that for
$\epsilon\rightarrow 0$ the commutator function
$G^+_{\text{AdS}}-G^-_{\text{AdS}}$ vanishes for
$\vert\eta^{\text{AdS}}\vert>1$. This result is in agreement with the
region where the retarded AdS Green function vanishes\footnote{For
values $\vert\eta^{\text{AdS}}\vert<1$ the hypergeometric function in
the expression for $G^\pm_{\text{AdS}}$ is defined via its analytic
continuation. This analytic continuation does depend on whether or not
the function depends on $\eta^{\text{AdS}}_{-\epsilon}$ or on
$\eta^{\text{AdS}}_{+\epsilon}$.} \cite{Danielsson:1998wt}.

Any two points $x$ and $x'$ in AdS for which $\eta^{\text{AdS}}(x,x')>1$ are
spacelike separated. Any two points $x$ and $x'\neq\bar x$ for which
$\eta^{\text{AdS}}(x,x')\le-1$ cannot be connected by any geodesic. The fact that
the commutator function vanishes for spacelike separated points is
often referred to as microcausality. The fact that the commutator
function also vanishes for points $x$ and $x'$ for which
$\eta^{\text{AdS}}(x,x')\le-1$ follows from the fact that the commutator function
(in the case of a free theory) is a classical object which must vanish
for points that cannot be connected by a classical path of
propagation. The commutator function is a continuous function of
$\eta^{\text{AdS}}$ and since it vanishes for $\eta^{\text{AdS}}>1$ it also vanishes for
$\eta^{\text{AdS}}=1$. Points $x$ and $x'$ for which $\eta^{\text{AdS}}(x,x')=1$ are separated
by a null geodesic. It turns out that in AdS all null geodesics are
also lightlike lines, that is achronal sets.

Summarising, we conclude that we have
\begin{equation}\label{eq:microcausality2}
 \lim_{\epsilon\to
0}\left(G^+_{\text{AdS}}-G^-_{\text{AdS}}\right)=0\hspace{1cm}\text{
for}\hspace{1cm}\vert\eta^{\text{AdS}}\vert\ge 1\,.
\end{equation}

\rnc{\Large}{\normalsize}

\end{document}